\documentclass[preprint,12pt]{elsarticle}

\usepackage[top=1truein,bottom=1truein,left=1truein,right=1truein]{geometry}

\graphicspath{{./pic}}

\usepackage{amsmath}
\usepackage{amssymb}
\usepackage{amsfonts}

\usepackage{hyperref}
\usepackage{url}
\usepackage{orcidlink}

\usepackage{here}
\usepackage{tikz}
\usepackage{bm}
\usepackage{setspace}
\usepackage{enumerate}
\usepackage{arydshln}
\usepackage{caption}
\usepackage{color}
\usepackage{appendix}
\usepackage{lineno}
\usepackage{ulem}
\usepackage{cancel}

\hypersetup{%
  colorlinks=true,
  linkcolor=magenta,
  citecolor=blue,
  urlcolor=cyan
}

\newcommand{\Jump}[1]{\ensuremath{[\![#1]\!]}}

\journal{Elsevier }

\begin{document}


\begin{frontmatter}

\title{Diffuse interface approach to oxygen transport and metabolism under blood flow dynamics in microcirculations}

\author[ad1]{Naoki Takeishi}
\author[ad2]{Junya Kobayashi}
\author[ad2]{Shigeo Wada}
\author[ad3]{Satoshi Ii\corref{cor}}

\address[ad1]{Department of Mechanical Engineering, Faculty of Engineering, Kyushu University, 744 Motooka, Nishi-ku, Fukuoka, 819-0395, Japan}
\address[ad2]{Graduate School of Engineering Science, The University of Osaka, 1-3 Machikaneyama Toyonaka, Osaka, 560-8531, Japan.}
\address[ad3]{Department of Mechanical Engineering, School of Engineering, Institute of Science Tokyo, Meguro-ku, 152-8550, Tokyo, Japan}

\cortext[cor]{Corresponding author: Satoshi Ii. Department of Mechanical Engineering, School of Engineering, Institute of Science Tokyo, Meguro-ku, Tokyo, 152-8550, Japan. \it{ii.s.148c@m.isct.ac.jp}}

\begin{abstract}
The relationship between the spatiotemporal distribution of oxygen transport and blood flow dynamics, accounting for the motion and deformation of individual red blood cells (RBCs), is of fundamental importance for understanding microcirculation systems. Three-dimensional (3D) modeling is indispensable for addressing complex oxygen transport and cellular behaviors in capillary networks; however, the computational approach is formidable for enforcing interface (or jump) conditions on largely moving and deforming interfaces. In this paper, we propose a diffuse interface approach for the oxygen transport using a mixture formulation. We formulate oxygen transport using an advection-diffusion-reaction equation and rewrite all governing equations in mixture forms using phase indicator functions, where all the interface conditions are included in the governing equations. This innovation avoids the complexity associated with discontinuities for largely moving interfaces in highly dense RBC conditions. We model cellular flow as a fluid-membrane interaction problem using the immersed boundary method (IBM). The method allows the seamless calculation of coupling problems for cellular flows and oxygen transports in the cytoplasm (internal fluid) of the RBC, plasma (external fluid), and tissue regions using a fixed Cartesian coordinate mesh. The proposed method accurately captures the analytical solution for spherically symmetric diffusion, and successfully demonstrates oxygen transport in both straight capillaries and their networks. These rigorous analyses suggest that RBCs can autonomously regulate the oxygen supply to tissues in response to the local tissue oxygenation level, resulting in the establishment of homogeneous tissue oxygenation.
\end{abstract}



\end{frontmatter}


\section{Introduction}
Mass transfer across a permeable, moving interface is frequently encountered in biological phenomena. A typical example of intravital mass transfer is oxygen transport, where RBCs are primarily responsible for carrying oxygen from the lungs to various organs through microcirculations~\cite{Krogh1929, Ramsook2023}. During this process, molecular oxygen diffuses from the alveoli to the interior of RBCs, where it binds to hemoglobin (Hb), a protein found in RBCs. The reverse process, that is, oxygen diffusion through the RBC membrane, plasma, and ultimately into tissues, occurs within the capillaries of various organs. Although it is well known that Hb's affinity for oxygen strongly depends on the local oxygen tension $P_{O_2}$~\cite{Ramsook2023, Manohar1986}, the complete oxygen transport process remains unclear because of its complexity. This process involves chemical reactions, convection, diffusion, and metabolic consumption within tissues. Because organs constitute a specialized homeostatic oxygen-sensing system and because tissues are capable of detecting local oxygen tension, understanding the relationship between microcirculatory blood flow and oxygen tension is of fundamental importance in identifying the biophysical basis of several diseases related to cellular-level metabolic activity~\cite{Weir2005}.

Because of limitations in experimental observation techniques, researchers have conducted theoretical and computational studies to understand oxygen transport and the maintenance of functions in {\it in vivo} systems. Numerical analysis began with the work of Krogh~\cite{Krogh1919}, who assumed that a capillary exchanged oxygen or other solutes only with a surrounding finite cylindrical tissue region, without considering individual RBCs. Then, a model was developed for a single capillary tube to consider the intraluminal resistance of a capillary vessel and the effects of RBCs as an oxygen carriers~\cite{Hellums1977, Federspiel1986, Hellums1995, Wang1993, Eggleton2000, Lucker2015, Lucker2017}. Because the oxygen supply to tissue occurs primarily through passive transport, the geometrical features of microvascular networks are not only basic biological parameters that influence tissue oxygenation but also fundamentally important for understanding the cellular metabolism~\cite{Schaffer2006, Lecoq2011, Lyons2016, Moeini2018}. Researchers have proposed reduced-order modelings to study blood supply and oxygen delivery within microvascular networks constructed based on two-photon microscopy imaging data~\cite{Gagnon2016, Gould2017, Hartung2018, Sweeney2018}, and mathematical vasculatures~\cite{Goldman2000, Koppl2020}. Despite these efforts, much remains unknown, in particular about how the behavior of individual RBCs relates to oxygen metabolism in microvascular systems.

In addition to those theoretical and numerical studies, recent computer simulation techniques have been used successfully to investigate aspects of cell dynamics, such as velocity and deformation. These simulations have reproduced both single-cell and multi-cellular dynamics in straight capillaries (e.g.,~\cite{Takeishi2014, Takeishi2015, Ii2018}). More recently, simulations of cellular-scale blood flow in complex microvascular networks have provided insight into the autoregulation of local variations in the volume fraction of RBCs (the so-called hematocrit, $Hct$) within the microvascular system~\cite{Balogh2017a, Balogh2017b}, where changes in local flow distribution are highly correlated with interactions between RBCs at branching points and the deformability of the RBC plays an important role in the flow regulation in vascular networks. Although these studies suggest that individual RBC dynamics, including their complex interactions, cannot be ignored in oxygen transport, no techniques have yet been developed to quantify the relationship among RBC-flowing behaviors, blood flows, and oxygen metabolism.

To solve these problems, a numerical approach to address mass transfer in different domains (or phases) with its moving interfaces is required. In these problems, physical quantities are smooth in each phase and governed by partial differential equations (PDEs), whereas, in general, the quantities have discontinuity at the interface (termed jump) to satisfy interfacial conditions, which are physical and mathematical conditions that bridge the solutions between PDEs in different phases at the interface. Because of a complexity in highly moving and deforming RBCs, in this study, we focus on the immersed (or embedded/shifted/unfitted) boundary method, which uses a non-boundary-fitted representation of the phase interface. An important issue from a numerical standpoint is how the interfacial conditions and associated quantity jumps are enforced when solving the PDEs with a moving interface on the non-conforming mesh. Regarding this, two major approaches exist for the representation of the interface: sharp interface and diffuse interface.

Sharp interface methods use a rigorous interface and address jumps across the interface using any reconstruction for unknowns. The ghost fluid method has been used to address a temperature jump on a level set-based sharp interface for the multiphase incompressible Navier--Stokes equations with phase change~\cite{Gibou2007}, which was originally developed for solving Stefan problems~\cite{Gibou2002}. Immersed interface methods that originated in work by \cite{LeVeque1994} for elliptic equations have been applied for a semi-permeable membrane~\cite{Layton2006, Jayathilake2010a, Jayathilake2010b, Jayathilake2011}. In these methods, jump quantities are incorporated into finite difference discretizations through the Taylor series expansions that include jumps of arbitrary-order derivatives. Miyauchi et al. formulated the incorporation of jumps into a finite element method for mass transfer on a selective permeable membrane~\cite{Miyauchi2015}. Wang et al. proposed a lattice Boltzmann method equipped with the interfacial jumps in heat and mass transfer problems using extrapolation with one-side stencils ~\cite{Wang2017}. The method was applied for solving oxygen transport with RBCs~\cite{Amiri2023a, Amiri2023b}. Zhao and Yan proposed an enriched IBM, which satisfies interfacial jump conditions by adding the nodal degrees of freedom in cut elements~\cite{Zhao2022}. Henneaux et al. enforced jump conditions for compressible viscous phases using a geometrically unfitted extended discontinuous Galerkin method~\cite{Henneaux2023}. However, the sharp interface approach essentially relies on the reconstruction of unknowns using several stencils around the interface; thus, geometrically-complex interface makes the procedure formidable in three dimensions.

By contrast, diffuse interface methods approximate the interface that is diffusive in its normal direction artificially and allow field quantities to smoothly distribute across the interface in the transition layer. Huang et al. proposed a numerical method that relies on the IBM for diffusion equations with various interfacial jump conditions. In this method, a smoothly distributed interfacial flux is introduced and updated in time as an augmented unknown~\cite{Huang2009}. Gong et al. extended Huang's idea for addressing the moving interface and solved the oxygen transport from multiple deformable RBCs~\cite{Gong2014}. The method was further developed for mass transfer through porous biomembranes~\cite{Wang2014}. However, regarding the time-evolution of the interfacial flux unknown, it is likely to smear out in the long-term calculation for oxygen transports in capillary networks. Amiri and Zhang have proposed the immersed membrane method, which introduces a transition layer between two phases, and approximately incorporated membrane resistance into the diffusion coefficient with a smoothed delta function~\cite{Amiri2021}. However, it is not clear whether the method can be extended to general interfacial conditions. Reder et al. proposed viscous stress approximations in the phase-field method for two-phase flow based on mechanical jump conditions. They formulated the jump of velocity gradients on the interface using phase mixture calculus~\cite{Reder2024}. These methods are based on the phase mixture formulation, which introduces mixture quantities and the interfacial conditions are implicitly incorporated into the basic equations; therefore, it is easy to address the moving interface. As an alternative approach to the single mixture equation, a multi-fluid model has been also proposed for coupling with the Navier--Stokes and Darcy equations, using the diffuse interface~\cite{Stoter2017}. Although many efforts have been devoted to addressing interfacial problems, a practical method for oxygen transport with membrane dynamics in capillary networks has not been established yet.

We aim to develop a diffuse interface approach for oxygen transport involving interfacial jump conditions using a mixture formulation, and couple it with cellular flow dynamics in capillary networks. In this study, the mathematical model of oxygen transport is based on a system of advection-diffusion-reaction type equations~\cite{Lucker2015}, where oxygen concentration and saturation are introduced in the cytoplasm of RBCs, plasma and the surrounding tissue regions and coupled with the chemical reaction (oxygen release from Hb with RBCs) and metabolic consumption of oxygen in tissues. The proposed mixture formulation rewrites the set of PDEs given in each phase as a single mixture PDE in the entire domain using a phase average operation with phase indicator functions. In this process, the quantity jumps at the interface are represented by the mixture quantities and interfacial conditions are implicitly incorporated into the mixture PDE. This formulation allows the seamless treatment of mass transfer in the entire domain without any geometric operations, such as the local reconstruction of quantities to address interfacial jumps that results in an easy implementation of implicit time discretization. Moreover, the present formulation discretely satisfies mass conservation because the interfacial jump is included in the diffusion flux. The cellular flow is modeled as the membrane and fluid structure interaction problem by the IBM~\cite{Peskin2002}, where the internal fluid (cytoplasm) and external fluid (plasma) are modeled as an incompressible Newtonian fluid and the RBC membrane is modeled as a hyperelastic membrane. Thus, the proposed method allows the seamless calculation of coupling problems for cellular flows and oxygen transports using a fixed Cartesian coordinate mesh. Numerical validations are demonstrated for spherically symmetric diffusion and a moving interface, and oxygen transport is investigated in simple capillary networks. To the best of our knowledge, this is a first attempt at the fully 3D modeling of oxygen transport that includes both RBC dynamics and nonlinear chemical reactions in submillimeter and second scales. We identify the relationship between the RBC distribution that includes individual moving-deforming behavior and tissue oxygenation within microvascular networks.

The remainder of this paper is organized as follows: In \S2, we provide the mathematical model of the mixture formulation; in \S3, we provide the numerical methodology; in \S4, we present the numerical results and verification, and in \S5, we provide concluding remarks.

\section{Mixture modeling for oxygen and Hb transports using diffuse interfaces}
%
\subsection{Definitions of material phases}
\begin{figure}[htbp]
  \centering
    \includegraphics[clip,width=16cm]{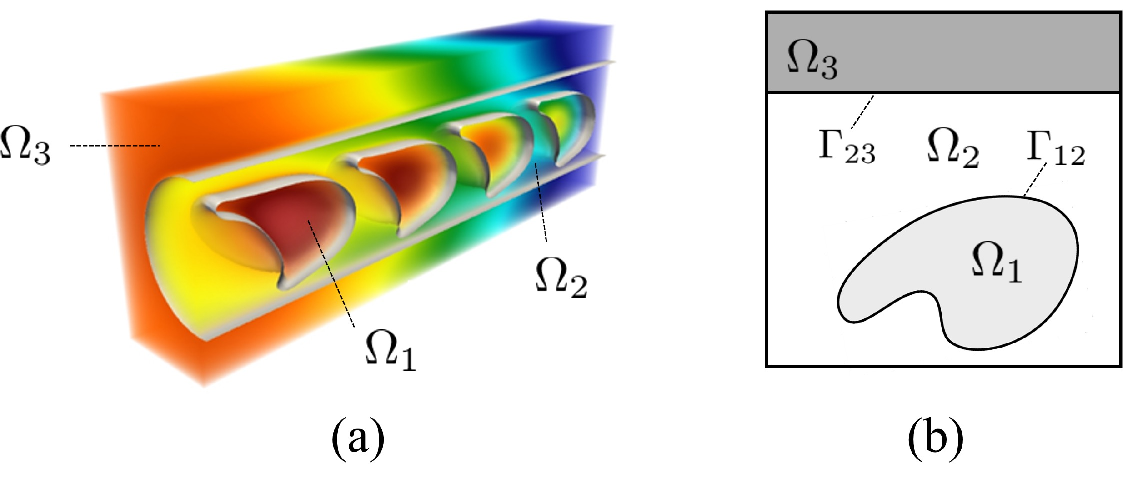}    
    \caption{
      (a) Schematic of the oxygen transport system in the present study. The contour colors represent the degree of oxygen concentration.
      (b) Definition of the computational regions $\Omega_i$ ($i \in [1,3]$), where $\Omega_1$ is the RBC interior (internal fluid) and $\Omega_2$ is plasma (external fluid), and their interfaces $\Gamma_{12}$ and $\Gamma_{23}$.
    }
    \label{fig:regions}
\end{figure}

We denote material phases by $\Omega_1$, $\Omega_2$ and $\Omega_3$ for the cytoplasm (internal fluid) of RBCs, plasma (external fluid) and tissue, and their interfaces by $\Gamma_{12} = \Omega_1 \cap \Omega_2$ and $\Gamma_{23} = \Omega_2 \cap \Omega_3$ for the cytoplasm-plasma and plasma-tissue interfaces, respectively (Figs.~\ref{fig:regions}(a) and \ref{fig:regions}(b)).
The spatial position and time are defined as ${\bf x} \in \Omega_1 \cup \Omega_2 \cup \Omega_3 \subset \mathbb{R}^3$ and $t\in[0,\cdot)\in\mathbb{R}$, respectively.

\subsection{Oxygen transport model}
A mathematical model of oxygen transport and consumption is based on existing models~\cite{Eggleton2000,Lucker2015}. The major features of the model are as follows:
\begin{itemize}
\item The amount of oxygen in each phase $\Omega_i$ ($i = 1, 2$ and $3$) is described by a macroscopic molar concentration $c_i({\bf x},t)$ [mlO$_2$/cm$^3$].
\item Oxygen concentration within the RBC is formulated in two states: the Hb associated with oxygen (oxygenation, Hb$\to$HbO$_2$) and the oxygenated Hb dissociated from oxygen (deoxygenation, HbO$_2$$\to$Hb), using the reaction rate $f_{Hb}$ [s$^{-1}$].
\item The oxygenated Hb is represented as oxygen saturation $s_{HbO_2}({\bf x},t) \in[0,1]$ in the internal RBC phase $\Omega_1$, which is given as a volume fraction of the Hb.
\item Oxygen is dissipated in the tissue phase to represent cell consumption with the metabolic rate of oxygen $M_c$ [mlO$_2$/(cm$^{3} \cdot$s)].
\item Oxygen concentration $c_i$ and saturation $s_{HbO_2}$ are governed by advection-diffusion-reaction type equations.
\item Oxygen concentration $c_i$ and partial pressure (oxygen tension) $P_i({\bf x},t)(=P_{O_2, i})$ [mmHg] are related by Henry's law.
\item The deformation of the RBC does not affect oxygen permeation across the membrane interface.
\end{itemize}

Oxygen transport and consumption are modeled in the entire region: $\Omega = \Omega_1 \cup \Omega_2 \cup \Omega_3$, where the oxygen is delivered by the RBC ($\Omega_1$), mainly diffused through plasma ($\Omega_2$), and finally released into tissue ($\Omega_3$) and consumed there. Dissolved oxygen within RBCs is quantified by its concentration $c_i$  and partial pressure $P_i$ for $\Omega_i$, which are related by Henry's law as
\begin{equation}
  c_i = \alpha_i P_i,
  \label{Henry}
\end{equation}
where $\alpha_i$ [mlO$_2$/(mmHg$\cdot$cm$^3$)] is the solubility coefficient given as $\alpha_1=3.38\times10^{-5}$, $\alpha_2=2.82\times10^{-5}$, and $\alpha_3=3.89\times10^{-5}$ mlO$_2$/(mmHg$\cdot$cm$^3$).

To describe the state of HbO$_2$, we consider the equilibrium oxygen saturation $s_{HbO_2}^{(eq)} = m_{HbO_2}/(m_{Hb} + m_{HbO_2})$ in $\Omega_1$, where $m_{HbO_2}$ and $m_{Hb}$ [mol/cm$^3$] are the amounts of substance of HbO$_2$ and Hb per unit volume, respectively.
Because $m$ and $P$ have a proportional relationship (i.e., $m \propto P$), $s_{HbO_2}^{(eq)}$ is redefined as the Hill equation:
\begin{equation}
  s_{HbO_2}^{(eq)} = \dfrac{P^n}{P_{(50)}^n + P^n}, 
  \quad {\rm for} \ {\bf x} \in \Omega_1,
  \label{s}
\end{equation}
where $P_{(50)}$ (= 47.9 mmHg) is oxygen tension at Hb half-saturation and $n$ (= 2.64) is the Hill exponent. We assume that Hb oxygenation is the positive reaction and the Hb deoxygenation is the negative reaction. This is formulated as
\begin{equation}
  \mathrm{Hb} + \mathrm{O}_2 \rightleftarrows \mathrm{HbO}_2.
  \label{Hb}
\end{equation}
The reaction rate of Hb (de)oxygenation $f_{Hb}$ can be described by assuming a non-equilibrium state:
\begin{equation}
  f_{Hb}(P,s_{HbO_2}) = 
  k_{off} \left( s_{HbO_2} - (1-s_{HbO_2}) \left( \dfrac{P}{P_{(50)}} \right)^n \right),
  \quad {\rm for} \ {\bf x} \in \Omega_1,
  \label{f}
\end{equation}
where $k_{off}$ [s$^{-1}$] is the dissociation rate set to $k_{off} = 44$ s$^{-1}$~\cite{Clark1985, Eggleton2000}. The Hb oxygenation state in $\Omega_1$ is characterized based on parameter $f_{Hb}$: equilibrium ($f_{Hb} = 0$, i.e., Eq.~\eqref{s} holds), Hb oxygenation ($f_{Hb} < 0$, i.e., from left to right in Eq.~\eqref{Hb}) and the Hb deoxygenation ($f_{Hb} > 0$, i.e., from right to left in Eq.~\eqref{Hb}). Therefore, the time derivation of $s_{HbO_2}$ without transport effects can be described using $f_{Hb}$ as
\begin{equation}
  \frac{\delta s_{HbO_2}}{\delta t} = -f_{Hb}(P,s_{HbO_2}).
\end{equation}

When oxygen consumption occurs only in tissue, the metabolic rate of oxygen, $M_c$, can be modeled using first-order Michaelis-Menten kinetics~\cite{Goldman2008}:
\begin{equation}
  M_c(P) = \dfrac{P}{P_{crit} + P} M_{\max}, 
  \quad {\rm for} \ {\bf x} \in \Omega_3,
\end{equation}
where $M_{\max}$ ($= 5$ $\mu$lO$_2$/(cm$^3 \cdot$s)) is the maximal metabolic rate of oxygen consumption and $P_{crit}$ ($= 1$ mmHg) is partial pressure at $M_c = M_{\max}/2$.

Oxygen concentration $c_i({\bf x},t)$ and saturation $s_1({\bf x},t) (=s_{HbO_2})$ are governed by advection-diffusion-reaction type equations:
\begin{subequations}
\begin{alignat}{2}
  & \frac{\partial c_i}{\partial t} + {\bf v} \cdot \nabla c_i = \nabla \cdot (D_i \nabla c_i) + e f_i - M_i,
  \quad & & {\rm for} \ {\bf x} \in \Omega_i, \quad (i=1,2,3),
  \label{adv_react_c} \\
  & \frac{\partial s_1}{\partial t} + {\bf v} \cdot \nabla s_1 = \nabla \cdot (D_{Hb} \nabla s_1) - f_1,
  \quad & & {\rm for} \ {\bf x} \in \Omega_1,
  \label{adv_react_s}
\end{alignat}
\end{subequations}
where ${\bf v}$ is the transport velocity (flow velocity); $e=N_{Hb}V_{mol}$ is the molar volume of oxygen, which is given by the product of the molar density $N_{Hb} = 2.03 \times 10^{-5}$ mol/cm$^3$ and molar volume of oxygen per mol $V_{mol} = 2.54 \times 10^4$ mLO$_2$/mol; $D_i$ is the diffusion coefficient of oxygen concentration in each phase given by $D_1=0.95\times10^{-7}$, $D_2=2.18\times10^{-7}$, and $D_3=2.41\times10^{-7}$ cm$^2$/s; and $D_{Hb}$ ($= 1.44 \times 10^{-7}$ cm$^2$/s) is the diffusivity of oxygenated Hb in RBC~\cite{Clark1985, Eggleton2000}. Note that $f_1=f_{Hb}(P_1,s_1)$, $f_2=f_3=0$, $M_1=M_2=0$, and $M_3=M_c(P_3)$.

According to the continuity of oxygen tensions (i.e., $P_1=P_2$, \cite{Eggleton2000}), the interfacial conditions for oxygen concentration on $\Gamma_{12}$ are given for the quantity:
\begin{equation}
  \frac{c_1}{\alpha_1} = \frac{c_2}{\alpha_2},
  \quad {\rm for} \ {\bf x} \in \Gamma_{12},
  \label{cont_c12}
\end{equation}
and the diffusion flux:
\begin{equation}
  D_1 \frac{\partial c_1}{\partial n} = D_2 \frac{\partial c_2}{\partial n},
  \quad {\rm for} \ {\bf x} \in \Gamma_{12},
  \label{flux_c12}
\end{equation}
where $\partial/\partial n (={\bf n}\cdot\nabla)$ is the normal derivative for the interface with the unit normal vector ${\bf n}$. Likewise, the interfacial conditions on $\Gamma_{23}$ are given by
\begin{subequations}
\begin{alignat}{2}
  &\frac{c_2}{\alpha_2}=\frac{c_3}{\alpha_3}, 
  \quad & & {\rm for} \ {\bf x} \in \Gamma_{23}, \\
  & D_2 \frac{\partial c_2}{\partial n} = D_3 \frac{\partial c_3}{\partial n}, 
  \quad & & {\rm for} \ {\bf x} \in \Gamma_{23}.
\end{alignat}
\end{subequations}
Since the oxygen saturation $s_1({\bf x},t) (=s_{HbO_2})$ can be defined within the cytoplasm $\Omega_1$ for physiological relevance,
reflecting the impermeability of (oxygenated) Hb due to its molecular size ($\approx 64$ kDa~\cite{Cooper2020}),
zero-flux condition is imposed on $\Gamma_{12}$ as the boundary condition:
\begin{equation}
  D_{Hb} \frac{\partial s_1}{\partial n} = 0,
  \quad {\rm for} \ {\bf x} \in \Gamma_{12}.
\end{equation}
%

\subsection{Definition of quantities in a mixture formulation}
To avoid the complexity of discontinuous problems across the phase interface, a set of mixture equations is derived from the aforementioned systems. We denote the jump across the interface $\Gamma_{12}$ from $\Omega_1$ to $\Omega_2$ by $\Jump{ \ }_{12}$ by
\begin{equation}
\Jump{ q ({\bf x}, t)}_{12} \equiv 
\lim_{\varepsilon \to 0} q_2 ({\bf x} + \varepsilon {\bf n}, t) 
- \lim_{\varepsilon \to 0} q_1 ({\bf x} - \varepsilon {\bf n}, t),
\quad \text{for} \ {\bf x} \in \Gamma_{12}.
\label{jump}
\end{equation}
In the mixture formulation, the jump $\Jump{q}_{12}$ is extended to the finite interface region $\tilde{\Gamma}_{12}$ described below and the interfacial conditions are redescribed by the extended jump $[q]_{12}$.

\begin{figure}[htbp]
  \centering
    \includegraphics[clip,width=16cm]{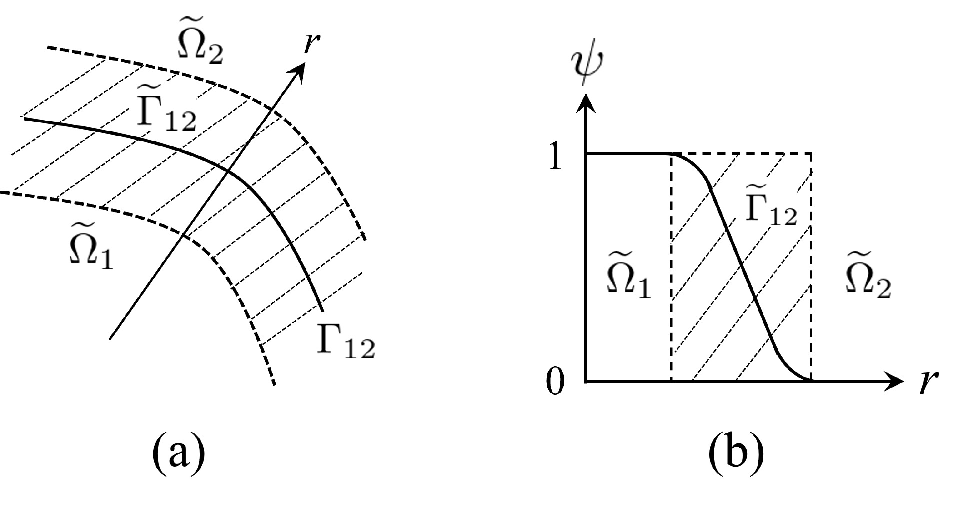}    
    \caption{
      (a) Schematic of an finite interface region $\tilde{\Gamma}_{12}$ (dashed area) representing the smoothed phase indicator function $\psi_1$ ($\in (0,1)$), and alternative phases: internal fluid $\tilde{\Omega}_1$ ($\psi_1 = 1$) and external fluid $\tilde{\Omega}_2$ ($\psi_1 = 0$).
      (b) Typical profile of the smoothed phase indicator function $\psi$ across the interface in the normal direction $r$.
    }
    \label{fig:mixture}
\end{figure}

As shown in Fig.~\ref{fig:mixture}, a finite interface region $\tilde{\Gamma}_{12}(t)$ is considered, and subsequently the alternative internal phase $\tilde{\Omega}_1(t)$ and external phase $\tilde{\Omega}_2(t)$ are redefined by the phase indicator function $\psi_1({\bf x}, t)$ as a smoothed phase indicator function:
\begin{equation}
  \begin{cases}
    \psi_1({\bf x},t) = 1, & \text{for} \ {\bf x} \in \tilde{\Omega}_1, \\
    0 < \psi_1({\bf x},t) < 1, & \text{for} \ {\bf x} \in \tilde{\Gamma}_{12}, \\
    \psi_1({\bf x},t) = 0, & \text{otherwise}.
  \end{cases}
  \label{psi_1}
\end{equation}
As the counterpart, the phase indicator function for $\tilde{\Omega}_2(t)$ is defined as $\psi_2({\bf x}, t) = 1 - \psi_1({\bf x}, t)$.

Using the phase indicator function $\psi_i ({\bf x}, t)$ and field variables $q_i({\bf x}, t)$ ($i=1,2$), a mixture quantity $q({\bf x}, t)$ and jump quantity $[q({\bf x}, t)]_{12}$ on $\tilde{\Gamma}_{12}$ are defined as
\begin{subequations}
\begin{align}
  & q = \psi_1 q_1 + \psi_2 q_2, \label{qa} \\
  & [q]_{12} = q_2 - q_1, \label{qJ12}
\end{align}
\end{subequations}
and vice versa,
\begin{subequations}
\begin{align}
  & q_1 = q - \psi_2[q]_{12}, \label{q1} \\
  & q_2 = q + \psi_1[q]_{12}. \label{q2}
\end{align}
\end{subequations}
Then, the spatial gradients of $q_1$ and $q_2$ are given by
\begin{subequations}
\begin{align}
  & \nabla{q_1} 
  = \nabla{q} - \nabla\psi_2[q]_{12} - \psi_2[\nabla{q}]_{12},
  \label{grad_q1} \\
  & \nabla{q_2} 
  = \nabla{q} + \nabla{\psi_1}[q]_{12} + \psi_1[\nabla{q}]_{12}.
  \label{grad_q2}
\end{align}
\end{subequations}
The gradient jump $[\nabla{q}]_{12}$ can be decomposed into normal and tangential components as
\begin{equation}
  [\nabla{q}]_{12} = {\bf n}[\partial_n{q}]_{12} + \nabla_s[{q}]_{12},
\label{jump_grad_q}
\end{equation}
where $\partial_n = \partial/\partial n$ is the normal derivative and $\nabla_s=({\bf I}-{\bf nn})\cdot\nabla$ is the surface (or tangential) derivative of the interface.
Note that the surface derivative is continuous in space; thus, $[\nabla_s{q}]_{12}=\nabla_s[q]_{12}$ holds.

\subsection{Derivation of mixture equations for oxygen concentration and oxygen saturation}
\subsubsection{Basic strategy}
In the present formulation, the field variables for $c({\bf x},t)$ and $s({\bf x},t)$ are described using the definitions Eqs.~\eqref{qa}--\eqref{q2}, and the mixture equations are derived from their mixture quantities. First, we start a formulation for binary phases and then extended it to three non-overlapping phases. In this study, we assume that the three phases are not a triplet formulation.

\subsubsection{Mixture equation for oxygen concentration for binary phases}
The continuity of oxygen tensions Eq.~\eqref{cont_c12} holds at the phase interface; thus, oxygen concentrations are discontinuous at the interface under different solubility coefficients $\alpha_1$ and $\alpha_2$. The interfacial condition for the continuity of oxygen tension with Henry's law can lead to a relationship between the material-phase variables and jump at the interface as follows:
\begin{equation}
  [c]_{12} = c_2 - c_1
  = \frac{\alpha_2}{\alpha_1}c_1 - c_1
  = \frac{[\alpha]_{12}}{\alpha_1} c_1
  = \frac{[\alpha]_{12}}{\alpha_2} c_2,
\end{equation}
where $[\alpha]_{12} = \alpha_2 - \alpha_1$. This can lead to
\begin{equation}
  c_1 = \frac{\alpha_1}{[\alpha]_{12}} [c]_{12}, \quad
  c_2 = \frac{\alpha_2}{[\alpha]_{12}} [c]_{12}.
  \label{c_12}
\end{equation}
Using the relation~\eqref{c_12}, Eq.~\eqref{qa} for $c$ can be written as
\begin{equation}
  c = \psi_1c_1 + \psi_2c_2 
  = \frac{\alpha}{[\alpha]_{12}}[c]_{12},
  \label{linear_Jc}
\end{equation}
where $\alpha=\psi_1\alpha_1+\psi_2\alpha_2$ is the mixture quantity for the solubility coefficient.
Thus, the jump $[c]_{12}$ can be given by the mixture variable $c$ as
\begin{equation}
  [c]_{12} = \frac{[\alpha]_{12}}{\alpha} c.
  \label{jump_c12}
\end{equation}
Finally, the material-phase concentration is simply expressed through Eq.~\eqref{c_12} as
\begin{equation}
  c_i = \frac{\alpha_i}{\alpha} c, \quad (i = 1, 2).
  \label{c_i}
\end{equation}

Using Eqs.~\eqref{grad_q1} and \eqref{grad_q2}, the continuity of the diffusion flux~\eqref{flux_c12} can be written as
\begin{equation}
  \begin{split}
    & D_2{\bf n}\cdot\nabla{c_2} - D_1{\bf n}\cdot\nabla{c_1} \\
    & = D_2{\bf n}\cdot(\nabla{c}+\nabla{\psi_1}[c]_{12}+\psi_1[\nabla{c}]_{12})
    - D_1{\bf n}\cdot(\nabla{c}-\nabla\psi_2[c]_{12}-\psi_2[\nabla{c}]_{12}) \\
    & = [D]_{12}{\bf n}\cdot\nabla{c}
    + (D_2 \partial_n{\psi_1} + D_1 \partial_n\psi_2)[c]_{12}
    + (D_2\psi_1 + D_1\psi_2)[\partial_n{c}]_{12} \\
    & = [D]_{12}{\bf n}\cdot\nabla{c}
    + [D]_{12}\partial_n{\psi_1}[c]_{12}
    + \hat{D}[\partial_n{c}]_{12} 
    \quad (\because \ \psi_2=1-\psi_1) \\
    & =0,
  \end{split}
\end{equation}
where $\hat{D}=\psi_2D_1+\psi_1D_2$ and $[D]_{12}=D_2-D_1$. This can lead to the normal gradient jump of $c$ as
\begin{equation}
  [\partial_n{c}]_{12} = 
  - \frac{[D]_{12}}{\hat{D}}
    \left( {\bf n}\cdot\nabla{c} + \partial_n{\psi_1}[c]_{12} \right).
  \label{jump_gradN_c}
\end{equation}
Thus, $[\nabla{c}]_{12}$ is written as
\begin{equation}
  \begin{split}
    [\nabla{c}]_{12}
    & = {\bf n}[\partial_n{c}]_{12} + \nabla_s[c]_{12} \\
    & = -{\bf n}\frac{[D]_{12}}{\hat{D}} 
    \left( {\bf n}\cdot\nabla{c} + \partial_n{\psi_1}[c]_{12} \right) 
    + \nabla_s[c]_{12} \\
    & = -\frac{[D]_{12}}{\hat{D}}{\bf n}{\bf n}\cdot\nabla{c} 
    -\frac{[D]_{12}}{\hat{D}}\nabla{\psi_1}[c]_{12}
    + \nabla_s[c]_{12} \\
    & = -\frac{[D]_{12}}{\hat{D}}\nabla{c} 
    -\frac{[D]_{12}}{\hat{D}}\frac{[\alpha]_{12}}{\alpha}\nabla{\psi_1}c
    + \left(\frac{[D]_{12}}{\hat{D}} 
    + \frac{[\alpha]_{12}}{\alpha} \right) \nabla_s{c},
  \end{split}
  \label{jump_grad_c}
\end{equation}
where $\nabla{\psi}_1={\bf n}\partial_n\psi_1+\cancelto{0}{\nabla_s\psi_1} \quad$ and the relation between the jump and mixture quantity~\eqref{jump_c12} are applied.

${\bf h}_i=D_i\nabla{c}_i$ ($i=1,2$) is introduced as a parameter related to the diffusion flux. Hence, the mixture form of the diffusion term is given by
\begin{equation}
  \begin{split}
  & \psi_1\nabla\cdot{\bf h}_1 + \psi_2\nabla\cdot{\bf h}_2 \\
  & \ = \nabla\cdot(\psi_1{\bf h}_1+\psi_2{\bf h}_2)
  -\nabla{\psi_1}\cdot{\bf h}_1 - \nabla{\psi_2}\cdot{\bf h}_2 \\
  & \ = \nabla\cdot{\bf h} 
  + \nabla{\psi}_1\cdot({\bf h}_2-{\bf h}_1) \\
  & \ = \nabla\cdot{\bf h} 
  + \partial_n{\psi}_1 \cancelto{0}{[{\bf n}\cdot{\bf h}]_{12}} \\
  & \ = \nabla\cdot{\bf h},
\end{split}
\end{equation}
where $[{\bf n}\cdot{\bf h}]_{12}=[D\partial_n{c}]_{12}=0$ holds from the flux continuity at the interface~\eqref{flux_c12}.

Using the jump conditions~\eqref{jump_c12} and \eqref{jump_grad_c}, the mixture quantity for the diffusion flux, ${\bf h}=\psi_1{\bf h}_1+\psi_2{\bf h}_2$, can be written as
\begin{equation}
  \begin{split}
    {\bf h} 
    & = \psi_1D_1\nabla{c_1} + \psi_2D_2\nabla{c_2} \\
    & = \psi_1 D_1 (\nabla{c}-\nabla\psi_2[c]_{12}-\psi_2[\nabla{c}]_{12})
    + \psi_2 D_2 (\nabla{c}+\nabla{\psi_1}[c]_{12}+\psi_1[\nabla{c}]_{12}) \\
    & = D\nabla{c}
    + D \nabla\psi_1[c]_{12}
    + \psi_1\psi_2 [D]_{12} [\nabla{c}]_{12} \\
    & = D^h\nabla{c}
    + D^h\frac{[\alpha]_{12}}{\alpha}\nabla\psi_1 c
    + \psi_1\psi_2 [D]_{12}\left(\frac{[D]_{12}}{\hat{D}} 
    + \frac{[\alpha]_{12}}{\alpha} \right) \nabla_s{c},
  \end{split}
  \label{mixture_f_acc}
\end{equation}
where $D=\psi_1D_1+\psi_2D_2$ is the mixture diffusion coefficient and $D^h$ denotes the mixture diffusion coefficient with the harmonic average:
\begin{equation}
  \begin{split}
    D^h 
    & = D - \psi_1\psi_2 \frac{[D]_{12}^2}{\hat{D}} \\
    & = \frac{(\psi_1D_1+\psi_2D_2)(\psi_2D_1+\psi_1D_2) - \psi_1\psi_2(D_2-D_1)^2}{\psi_2D_1+\psi_1D_2} \\
    & = \frac{(\psi_1\psi_2D_1^2+\psi_2^2D_1D_2+\psi_1^2D_1D_2+\psi_1\psi_2D_2^2) - \psi_1\psi_2(D_1^2+D_2^2-2D_1D_2)}{\psi_2D_1+\psi_1D_2} \\
    & = \frac{D_1D_2}{\psi_2D_1+\psi_1D_2}
    \left(= \frac{1}{\psi_1/D_1+\psi_2/D_2}\right).
  \end{split}
  \label{Dh}
\end{equation}
Obviously, from~\eqref{mixture_f_acc}, anisotropic diffusion occurs ${\bf D}^{a}\cdot\nabla{c}$ with the anisotropic diffusion coefficient tensor ${\bf D}^{a}$ consisting of the normal and tangential components:
\begin{align}
    {\bf D}^{a} 
    & = D^h{\bf I} + \psi_1\psi_2 [D]_{12}\left(\frac{[D]_{12}}{\hat{D}} + \frac{[\alpha]_{12}}{\alpha} \right) ({\bf I}-{\bf nn}), \\
    & = D{\bf I} 
    - \psi_1\psi_2\frac{[D]_{12}^2}{\hat{D}}{\bf nn}
    + \psi_1\psi_2\frac{[D]_{12}[\alpha]_{12}}{\alpha}({\bf I}-{\bf nn}), \label{Da_2}
\end{align}
where the second and third terms of the right-hand side of Eq.~\eqref{Da_2} denote the components for derivatives in the normal and tangential directions, respectively, which only act on the smooth interface region $\tilde{\Gamma}_{12}$ because $\psi_1\psi_2=\psi_1(1-\psi_1)$ is not zero in this region. In our application of oxygen transport across the RBC membrane, accurate treatment in the normal direction is more important than in the tangential direction. Determining a numerical treatment for anisotropic diffusion is challenging because discrete stencils increase, which results in a larger computational cost than the isotropic case. For instance, if a second-order central difference method is applied, the local stencil becomes $3^3=27$ in 3D which is much larger than $7$ for the discretization of the isotropic diffusion problem. Thus, in this study, we make the following assumption:
\begin{equation}
  {\bf D}^{a} \approx D^h{\bf I},
  \label{eq:Da_approx}
\end{equation}
which therefore approximates Eq.~\eqref{mixture_f_acc} as
\begin{equation}
  {\bf h} \approx
  D^h\nabla{c} + D^h\frac{[\alpha]_{12}}{\alpha}\nabla\psi_1 c.
\end{equation}
Note that a similar form for the anisotropic transfer was derived in a different diffuse interface approach using tensorial mobilities~\cite{Nicoli2011}.

\subsubsection{Extension to three non-overlapping three phases}
We assume that the direct interaction between RBCs and tissue can be ignored; therefore, interfacial conditions are only considered for $\Gamma_{12}$ and $\Gamma_{32}$.
Under this assumption, the mixture model can be easily extended to three phases.
We introduce a smoothed phase indicator function $\psi_3({\bf x},t)$, independently of $\psi_1({\bf x},t)$~\eqref{psi_1}, as 
\begin{equation}
  \begin{cases}
    \psi_3({\bf x},t) = 1, & \text{for} \ {\bf x} \in \tilde{\Omega}_3(t), \\
    0 < \psi_3({\bf x},t) < 1, & \text{for} \ {\bf x} \in \tilde{\Gamma}_{32}(t), \\
    \psi_3({\bf x},t) = 0, & \text{otherwise}.
  \end{cases}
  \label{psi_3}
\end{equation}
Then the phase mixture is redefined as $q=\psi_1q_1+\psi_2q_2+\psi_3q_3$, where $\psi_2=1-\psi_1-\psi_3$, and the jump on $\tilde{\Gamma}_{32}$ is defined as $[q]_{32}=q_2-q_3$.

According to the aforementioned assumption, we conclude that our mixture system can be written as
\begin{equation}
  \frac{\partial c}{\partial t} + {\bf v} \cdot \nabla c 
  = \nabla \cdot \left( D^h \nabla c + \frac{D^h}{\alpha} \left( [\alpha]_{12} \nabla \psi_1 + [\alpha]_{32} \nabla \psi_3 \right) c
    \right) 
  + \psi_1 e f_1 - \psi_3 M_3.
  \label{mixture_system}
\end{equation}
%

\subsubsection{Mixture equation for the transport of oxygen saturation}
First we rewrite the diffusion term with the interfacial jump in Eq.~\eqref{mixture_system} for oxygen saturation $s$ as
\begin{equation}
  D^h \nabla c + \frac{D^h}{\alpha} 
  \left( [\alpha]_{12} \nabla \psi_1 + [\alpha]_{32} \nabla \psi_3 \right) c 
  \Rightarrow
  D^h_s \nabla s + \frac{D^h_s}{\beta} 
  \left([\beta]_{12} \nabla \psi_1 + [\beta]_{32} \nabla \psi_3 \right) s,
  \label{Hb_saturation_s}
\end{equation}
where $D^h_s$ and $\beta$ correspond to the diffusion coefficient $D^h$ and solubility coefficient $\alpha$ for $c$, respectively. In nature, Hb molecules are not defined in $\Omega_2$ and $\Omega_3$; thus, the governing equation for oxygen saturation is not necessary in these regions.
However, we apply the mixture formulation for the solution algorithm on fixed meshes; thus, the governing equation must be extended to these domains.

By introducing non-physical parameters $D_{s2}=D_{s3}=\varepsilon_{D}D_{Hb}$, we can provide the following derivation:
\begin{equation}
  D^h_s 
  = \frac{1}{\psi_1/D_{Hb}+\psi_2/D_{s2}+\psi_3/D_{s3}}
  = \frac{D_{Hb}}{\psi_1+(1-\psi_1)/\varepsilon_{D}}.
  \label{Dsh}
\end{equation}
Analogously, by introducing non-physical parameters $\beta_1=1$ and $\beta_2=\beta_3=\varepsilon_{\beta}$, we obtain
\begin{equation}
  \beta 
  = \beta_1\psi_1 +\beta_2\psi_2 +\beta_3\psi_3
  = \psi_1 + (1-\psi_1)\varepsilon_{\beta}
\end{equation}
and
\begin{equation}
  [\beta]_{12} = \beta_2 - \beta_1 = \varepsilon_{\beta} - 1, 
  \quad {\rm and} \quad [\beta]_{32} = 0.
\end{equation}
Thus, the diffusion flux~\eqref{Hb_saturation_s} is rewritten as
\begin{equation}
  D_s^h \left(\nabla s +\frac{\varepsilon_{\beta}-1}{\psi_1 + (1-\psi_1)\varepsilon_{\beta}} \nabla\psi_1 s \right).
\end{equation}

The above derivation can be obtained by reconsidering the following interfacial conditions:
\begin{subequations}
\begin{align}
  &\begin{cases}
  \varepsilon_{\beta} s_1 = s_2,
  & {\rm for} \ {\bf x}\in\Gamma_{12}, \\
  \varepsilon_{D} \partial_n s_1 = \partial_n s_2, 
  & {\rm for} \ {\bf x}\in\Gamma_{12},
  \end{cases} \\
  &\begin{cases}
  s_2 = s_3, 
  & {\rm for} \ {\bf x}\in\Gamma_{23} \\
  \partial_n s_2 = \partial_n s_3,
  & {\rm for} \ {\bf x}\in\Gamma_{23}.
  \end{cases}
\end{align}
\end{subequations}
Therefore, if the parameters $\varepsilon_{\beta}$ and $\varepsilon_{D}$ are sufficiently small ($\ll$ 1), the artificial oxygen saturation $s_2$ and $s_3$ converge to zero constant asymptotically. In this study, we set $\varepsilon_{\beta}=10^{-5}$ and $\varepsilon_{D} = 4 \times 10^{-3}$ from several numerical trials to obtain stable numerical solutions, which causes a small amount of leakage of oxygen saturation from the RBC because of numerical dissipation. We refer to this point in \S 4 through numerical examples.

\section{Numerical methods}
\subsection{Modeling of cell suspension flows}
\subsubsection{Coupling between the fluid and elastic membrane}
The fluid behaviors in both $\Omega_1$ and $\Omega_2$ are governed by the continuity and incompressible Navier-Stokes equations:
\begin{subequations}
\begin{align}
  & \nabla \cdot {\bf v} = 0, \\
  & \rho \left( \frac{\partial{\bf v}}{\partial t} 
  + {\bf v} \cdot \nabla {\bf v} \right) 
  = -\nabla{p} + \nabla\cdot{\bm \tau} + {\bf F}, 
  \label{ns_equation} \\
  & {\bm \tau} = \mu \left( \nabla{\bf v} + \nabla {\bf v}^T \right),
\end{align}
\end{subequations}
for ${\bf x}=(x, y, z) \in \Omega_1\cup\Omega_2$, where $\rho$ is the fluid density, ${\bf v}({\bf x}, t)$ is the velocity vector, $p({\bf x}, t)$ is the pressure, ${\bf F}({\bf x},t)$ is the membrane force, and ${\bm \tau}({\bf x}, t)$ is the viscous stress tensor with the mixture viscosity $\mu=\psi_1\mu_1 + \psi_2\mu_2$.

The IBM~\cite{Peskin2002} is coupled with the fluid and membrane mechanics.
The force density vector ${\bf f}_m(t)={\bf f}({\bf x}_m,t)$ at the membrane node ${\bf x}_m(t)$ is distributed to the neighboring fluid position ${\bf x}$, and the external force ${\bf F}({\bf x},t)$ in Eq.~\eqref{ns_equation} is described as 
\begin{equation}
  {\bf F} ({\bf x},t) = \int_{S} \mathcal{D}({\bf x}-{\bf x}_m(t)) {\bf f}_m(t) dS,
  \label{force}
\end{equation}
where $S=\Gamma_{12}$, and $\mathcal{D}({\bf x})$ is a smoothed delta function approximating the Dirac delta function. In this study, we use the approximate function used in~\cite{Peskin2002}:
\begin{equation}
  \mathcal{D}({\bf x}) =
  \begin{cases}
    \dfrac{1}{8 \Delta x^3} \prod_{j=1}^{3} \left( 3 - 2|x_j| + \sqrt{1 + 4|x_j| - 4 |x_j|^2} \right), 
    & \text{for} \ \ 0 \leq |x_j| < \Delta x, \\
    \dfrac{1}{8 \Delta x^3} \prod_{j=1}^{3} \left(5 - 2|x_j| - \sqrt{-7 + 12|x_j| - 4|x_j|^2} \right),
    & \text{for} \ \ \Delta x \leq |x_j| < 2\Delta x, \\
    0, 
    & \text{for} \ \ 2\Delta x \leq |x_j|,
  \end{cases}
  \label{Dirac-Delta}
\end{equation}
where $\Delta x$ is the lattice size, and $x_1 = x$, $x_2 = y$, and $x_3 = z$, respectively. The velocity at the membrane node ${\bf v}_m(t)={\bf v} ({\bf x}_m,t)$ is obtained by interpolating the velocities at fluid nodes:
\begin{equation}
  {\bf v}_m(t) = \int_{\Omega} \mathcal{D}({\bf x} - {\bf x}_m(t)) {\bf v} ({\bf x},t) d{\bf x}.
  \label{velocity}
\end{equation}
The membrane node point ${\bf x}_m(t)$ is assumed to follow the background velocity field.
Thus, it is updated using Lagrangian tracking with the no-slip condition:
\begin{equation}
  \frac{d {\bf x}_m}{dt} = {\bf v}_m.
  \label{lagrangian}
\end{equation}
We have confirmed that this assumption is reasonable,
at least for the validation of deformation index of single RBC and for cell-depleted peripheral layer thickness in the multi-RBC interaction problem (Figure A1C and A1D in~\cite{Takeishi2014}).

\subsubsection{Membrane model}
The membrane force is composed of the in-plane force ${\bf f}_s$ and bending force ${\bf f}_b$:
\begin{equation}
  {\bf f} = {\bf f}_s + {\bf f}_b.
\end{equation}

The in-plane stress is modeled as an isotropic and hyperelastic material. The surface deformation gradient tensor ${\bf F}_s$ is given by
\begin{equation}
  d{\bf x}_m = {\bf F}_s \cdot d{\bf X}_m,
  \label{Fs}
\end{equation}
where ${\bf X}_m$ and ${\bf x}_m$ are the membrane positions of the reference and deformed states, respectively. The local deformation of the membrane can be measured using the surface Green-Lagrange strain tensor:
\begin{equation}
  {\bf E}_s = \frac{1}{2} \left( {\bf C}_s - {\bf I}_s \right),
  \label{e}
\end{equation}
where ${\bf C}_s = {\bf F}_s^T\cdot{\bf F}_s$ is the right surface Cauchy-Green tensor and ${\bf I}_s$ $(= {\bf I} - {\bf n}{\bf n}$) is the tangential (or surface) projection operator with the outward unit normal vector of the interface, ${\bf n}$.
The two invariants of the in-plane strain tensor ${\bf E}_s$ can be given by 
\begin{equation}
  I_1 = \eta_1^2 + \eta_2^2 - 2, \quad I_2 = \eta_1^2 \eta_2^2 - 1 = J_s^2 - 1,
  \label{I1andI2}
\end{equation}
where $\eta_1$ and $\eta_2$ are the principal extension ratios. The Jacobian $J_s = \eta_1\eta_2$ expresses the ratio of the deformed surface area to the reference surface area.
The elastic stresses in an infinitely thin membrane are replaced by elastic tensions.
The Cauchy stress tensor (in-plane stress tensor) ${\bf T}_s$ can be related to an elastic strain energy per unit area $w_s \left( I_1, I_2 \right)$:
\begin{equation}
  {\bf T}_s = \frac{1}{J_s} {\bf F}_s \cdot \frac{\partial w_s \left( I_1, I_2 \right)}{\partial {\bf E}_s} \cdot {\bf F}_s^T,
  \label{T_s}
\end{equation}
where $w_s=w_s^{SK}$ satisfies the SK law~\cite{Skalak1973}:
\begin{equation}
  w_s^{SK} \left( I_1, I_2 \right) 
  = \frac{G_s}{4} \left( I_1^2 + 2 I_1 - 2I_2 + C I_2^2\right),
  \label{SK}
\end{equation}
where $G_s$ is the surface elastic modulus and $C$ is the area dilation modulus.
In this study, we set $G_s$ = 5 $\mu$N/m and $C = 50$.

Neglecting inertial effects on membrane deformation, the static local equilibrium equation of the membrane is given by 
\begin{equation}
  \nabla_s \cdot {\bf T}_s + {\bf f}_s = {\bf 0}.
  \label{StrongForm}
\end{equation}
Based on the virtual work principle, the above equation in strong form~\eqref{StrongForm} can be rewritten in weak form as 
\begin{equation}
  \int_S \hat{{\bf u}} \cdot {\bf f}_s dS = \int_S \hat{\boldsymbol{\epsilon}} : {\bf T}_s dS,
  \label{WeakForm}
\end{equation}
where $\hat{{\bf u}}$ and $\hat{\boldsymbol{\epsilon}} = (\nabla_s \hat{{\bf u}} + \nabla_s \hat{{\bf u}}^T )\big/2$ are the virtual displacement and virtual strain, respectively.
The finite element method with a linear (triangular) element is used to solve Eq.~\eqref{WeakForm}. In this study, to reduce computational cost, we adopt a mass-lumped technique for solving the linear system~\eqref{WeakForm}, in which the coefficient matrix in the linear system becomes diagonal and easy to solve.

The membrane bending force ${\bf f}_b$ is given by the Helfrich model~\cite{Zhongcan1989} as
\begin{equation}
  {\bf f}_b = E_b \left(
    (2\kappa_m+c_0)(2\kappa_m^2 -2\kappa_g -c_0\kappa_m) 
    + 2\Delta_{LB}\kappa_m \right) {\bf n},
  \label{bending_fb}
\end{equation}
where $E_b$ is the bending modulus, $c_0$ is the spontaneous curvature, $\kappa_m$ and $\kappa_g$ are the mean and Gaussian curvatures, respectively, and $\Delta_{LB}=\nabla_s\cdot\nabla_s$ is the Laplace-Beltrami operator. In this study, we set $E_b$ = 1.8$\times$10$^{-19}$ J and $c_0=0$. We apply the discrete divergence theorem-based method~\cite{Xu2004,Reuter2009} for calculating $\Delta_{LB}\kappa_m = \nabla_s\cdot(\nabla_s\kappa_m)$, as shown in a previous study~\cite{Yazdani2012}. We also apply a similar discretization for calculating the curvature tensor ${\bm \kappa}=\nabla_s{\bf n}$. The unit normal vectors ${\bf n}$ are calculated at the element nodes, which are averaged by the normal vectors on neighboring elements. Once the curvature tensor is calculated, the mean and Gaussian curvatures are given as $\kappa_m=I_1({\bm \kappa})$ and ${\kappa}_g=I_2({\bm \kappa})$, where $I_1(\cdot)$ and $I_2(\cdot)$ are the first and second invariants, respectively.

\subsection{Calculation of the phase indicator function $\psi_1$}
Based on the front-tracking method~\cite{Unverdi1992}, the phase indicator function describing the RBC inside, $\tilde{\psi}_1$, is obtained by solving Poisson's equation:
\begin{equation}
  \nabla^2 \tilde{\psi}_1 
  = \nabla \cdot \left(\int_S \mathcal{D}({\bf x}-{\bf x}_m) {\bf n}_m dS \right),
  \label{poisson_psi}
\end{equation}
where $\tilde{\cdot}$ indicates the temporal quantity.
As we describe later, we update $\psi_1$ using the interface capturing method; thus, $\psi_1$ is slightly modified so that the value range is $\psi_1\in[0,1]$. To achieve this, we introduce a procedure in which the phase indicator function is recalculated from the signed distance function. We introduce a function, which maps the indicator function $\tilde{\psi}_1$ to the level-set function $\Theta_1$ for the signed distance from the interface $\Gamma_{12}$, as $\mathcal{G}: \tilde{\psi}_1 \mapsto \Theta_1$. For the mapping, we apply a method using multi-dimensional tangent of hyperbola for interface capturing (MTHINC) reconstruction~\cite{Ii2012, Ii2018}. Then, the phase indicator function is calculated using the smooth function $\mathcal{H}$:
\begin{equation}
  \mathcal{H} (\Theta^{\ast}) = 
  \begin{cases}
    0, & \ {\rm for} \ \Theta^{\ast}<-1, \\
    \frac{1}{2}\left(
      1 +\Theta^{\ast} +\frac{1}{\pi}\sin(\pi\Theta^{\ast})
    \right), &  \ {\rm for} \ -1 \leq \Theta^{\ast} \leq 1, \\
    1, & \ {\rm for} \ \Theta^{\ast} >1,
  \end{cases}
  \label{eq:psi_SDF}
\end{equation}
as $\psi_1=\mathcal{H}(\Theta_1^\ast)$, which can result in
\begin{equation}
  \psi_1 = \mathcal{H}(\mathcal{G}(\tilde{\psi}_1)),
\end{equation}
where $\Theta^\ast = \Theta/h = \Theta/\Delta x$.

Solving the Poisson equation~\eqref{poisson_psi} takes computational time; therefore, we solve the advection equation for updating $\psi_1$ by applying a less dissipative numerical method (i.e., interface capturing method) without the Poisson equation every time. The phase indicator function $\psi_1$ in the fluid phase ($\tilde{\Omega}_1 \cup \tilde{\Omega}_2$) is governed by an advection equation:
\begin{equation}
  \frac{\partial \psi_1}{\partial t} + \nabla \cdot \left( {\bf v} \psi_1 \right) - \psi_1 \nabla \cdot {\bf v} = 0.
  \label{VOF}
\end{equation}

\subsection{Fluid-rigid wall interaction}
In this study, tissue is assumed be rigid; thus, the Dirichlet boundary condition for the velocity is applied at the interface between plasma and tissue, $\Gamma_{23}$, that is, considering the fixed vessel wall.
Because oxygen transport is handled by the mixture formulation, we use the boundary data immersion (BDI) method~\cite{Weymouth2011}, which also applies the phase mixture formulation. The governing equations for the fluid and solid (or rigid) are given
\begin{equation}
  \begin{cases}
  \rho \left( \frac{\partial{\bf v}}{\partial t} 
  + {\bf v} \cdot \nabla {\bf v} \right) 
  = - \nabla p + \nabla\cdot{\bm \tau} + {\bf F},
  & \ \text{for} \ {\bf x}\in\Omega_1 \cup \Omega_2, \\
  {\bf v} = 0,
  & \ \text{for} \ {\bf x}\in\Omega_3.
  \end{cases}
\end{equation}
To apply the BDI method, the momentum equations are rewritten using integration during a time interval $\tau\in[t', t]$:
\begin{equation}
  \begin{cases}
  {\bf v}|_{t}
  = 
  {\bf v}|_{t'}
  + \int_{t'}^{t} 
  \left(
  - {\bf v} \cdot \nabla {\bf v}
  + \frac{1}{\rho}\left( 
    - \nabla p + \nabla\cdot{\bm \tau} + {\bf F}
     \right) \right) d\tau,
  & \ \text{for} \ {\bf x}\in\Omega_1 \cup \Omega_2, \\
  {\bf v}|_{t} = 0,
  & \ \text{for} \ {\bf x}\in\Omega_3,
  \end{cases}
\end{equation}
where $\cdot|_{t}$ denotes the quantity at time $t$. Considering the phase mixture average using the phase indicator function for the tissue phase, $\psi_3$, the above equations become
\begin{equation}
  {\bf v}|_{t}
  = (1-\psi_3)
  \left(
    {\bf v}|_{t'}
    + \int_{t'}^{t} 
    \left(
    - {\bf v} \cdot \nabla {\bf v}
    + \frac{1}{\rho}\left( 
      - \nabla p + \nabla\cdot{\bm \tau} + {\bf F}
      \right) \right) d\tau \right),
  \quad \text{for} \ {\bf x}\in\Omega.
\end{equation}
The solenoidal vector field of the velocity ${\bf v}|_{t}$ is achieved by satisfying the continuity condition:
\begin{equation}
 \nabla\cdot{\bf v}|_{t} = 0,
 \quad \text{for} \ {\bf x}\in\Omega.
\end{equation}

\subsection{Summary of the system}
The final system of mathematical models with the mixture formulation is given by
\begin{subequations}
\begin{align}
  & \nabla \cdot {\bf v} = 0, \label{mix_div_u} \\
  & {\bf v}|_{t}
  = 
  (1-\psi_3) 
  \left(
    {\bf v}|_{t'}
  + \int_{t'}^{t} 
  \left(
  - {\bf v} \cdot \nabla {\bf v}
  + \frac{1}{\rho}\left( 
    - \nabla p + \nabla\cdot{\bm \tau} + {\bf F}
     \right) \right) d\tau \right),
  \label{mix_moment} \\
  & {\bm \tau} 
  = \mu(\nabla{\bf v}+\nabla{\bf v}^T), \\
  & {\bf F} = \int_{S} \mathcal{D}({\bf x}-{\bf x}_m) {\bf f}_m dS, \\
  & \frac{d{\bf x}_m}{dt} 
  = \int_{\Omega} \mathcal{D}({\bf x}-{\bf x}_m) {\bf v} d{\bf x}, 
  \label{dxm}
  \\
  & \frac{\partial c}{\partial t} 
  + {\bf v} \cdot \nabla c 
  =
  \nabla \cdot \left(
     D^h \nabla c + \frac{D^h}{\alpha} \left( [\alpha]_{12} \nabla \psi_1+ [\alpha]_{32}\nabla \psi_3  \right) c
  \right)
  + ef - M,
  \label{mix_c}\\
  & \frac{\partial s}{\partial t} + {\bf v} \cdot \nabla s 
  = \nabla \cdot \left(
    D_s^h \nabla s +\frac{D_s^h(\varepsilon_{\beta}-1)}{\psi_1 + (1-\psi_1)\varepsilon_{\beta}} \nabla\psi_1 s
    \right) 
  - f,
  \label{mix_s} \\
  & \frac{\partial \psi_1}{\partial t} 
  + {\bf v}\cdot\nabla{\psi_1}
  = 0, \label{adv_psi} \\
  & \quad {\rm or} \quad
  \psi_1 = \mathcal{H}(\mathcal{G}(\tilde{\psi}_1)), 
  \ \ \nabla^2\tilde{\psi}_1 
  = \nabla \cdot \left(\int_S \mathcal{D}({\bf x}-{\bf x}_m) {\bf n}_m dS \right),
\end{align}
\end{subequations}
for ${\bf x}\in\Omega$, where $f = \psi_1 f_{Hb}(P_1,s_1)$ and $M = \psi_3 M_c(P_3)$ with
\begin{subequations}
\begin{align}
  & s_1 = \frac{\beta_1 s}{\beta} = \frac{s}{\psi_1 + (1-\psi_1)\varepsilon_{\beta}}, \\
  & P_1 = P_3 (= P_2) = \frac{c}{\alpha}.
\end{align}
\end{subequations}
Note that $\psi_3$ is not changed over time, that is, $\psi_3({\bf x},t)=\psi_3({\bf x})|_{t=0}$, and $\psi_2=1-\psi_1-\psi_3$.

\subsection{Discretizations}
For discretization, the Cartesian coordinate mesh is used and a conservative finite difference (or finite volume) method are used.

Oxygen concentration $c$ and saturation $s$ are solved using a fractional step method, which decomposes the equation into the advection parts:
\begin{subequations}
\begin{align}
  & \frac{\partial c}{\partial t} +{\bf v}\cdot\nabla{c} = 0, 
  \label{adv_c} \\
  & \frac{\partial s}{\partial t} +{\bf v}\cdot\nabla{s} = 0,
  \label{adv_s}
\end{align}
\end{subequations}
and non-advection parts:
\begin{subequations}
\begin{align}
  & \frac{\partial c}{\partial t}
  = \nabla\cdot \left(
     D^h\nabla{c} +\frac{D^h}{\alpha}
     \left([\alpha]_{12}\nabla\psi_1 +[\alpha]_{32}\nabla\psi_3  \right) c
  \right)
  + ef + M, \label{nonadv_c} \\
  & \frac{\partial s}{\partial t}
  = \nabla\cdot\left( 
    D_s^h \nabla s +\frac{D_s^h(\varepsilon_{\beta}-1)}{\psi_1 + (1-\psi_1)\varepsilon_{\beta}} \nabla\psi_1 s
    \right) 
  - f. \label{nonadv_s}
\end{align}
\end{subequations}
The fifth-order targeted ENO (TENO) scheme~\cite{Fu2016} is applied for solving the advection equations~\eqref{adv_c} and \eqref{adv_s}, and the second-order central difference method is applied for the non-advection terms in Eqs.~\eqref{nonadv_c} and \eqref{nonadv_s}. The explicit second-order Runge--Kutta method is applied for Eqs.~\eqref{adv_c} and \eqref{adv_s}, and the Crank--Nicolson method and explicit Euler method are used for the diffusion and source (reaction) terms, respectively, in Eqs.~\eqref{nonadv_c} and \eqref{nonadv_s}. To avoid numerical instability in the evaluation of the reaction terms, $s$ is enforced as bounded to $s\in[0,1]$ in all time integrations.

The fluid-capsule coupling system is solved using the projection-type method with a staggered arrangement of variables~\cite{Harlow1965}; thus, the fluid-rigid coupling equations obtained by the BDI method~\eqref{mix_div_u} and \eqref{mix_moment} are solved using the fractional step method~\cite{Ii2018b}. The convection terms are discretized using the fifth-order WENO method~\cite{Jiang1996} to avoid a numerical oscillation, whereas the viscous terms are discretized using the second-order central difference method. For temporal integrations, the second-order Adams--Bashforth method is applied to the convection terms and the Crank--Nicolson method is applied to the viscous terms. The explicit second-order Runge--Kutta method is applied to solve Eq.~\eqref{dxm}. The BiCGStab solver with geometrical multigrid pre-conditioning using the SOR solver is used to solve the linear systems for the prediction equations for the velocity, and Poisson equations for the pressure and phase indicator function.

Eq.~\eqref{adv_psi} is solved using the MTHINC method \cite{Ii2012}, in which a multi-dimensional continuous function is reconstructed to approximate the phase indicator function, originally developed by the THINC scheme~\cite{Xiao2005}. Linear interface reconstruction is applied. To avoid numerical inconsistency, $\psi_1$ is reinitialized by solving Eq.~\eqref{poisson_psi} once in a thousand steps.

In this study, capillary networks are given by considering Boolean operations for some cylinders (for the vessel) and spheres (for the junction); thus, theoretical expressions of the geometries are always provided. Therefore, the phase indicator function for the tissue phase, $\psi_3$, is evaluated by numerically integrating a (sharp) indicator function given by Eq.~\eqref{eq:psi_SDF} with $h=0$ in each grid cell. This indicates that $\psi_3$ is a volume-of-fluid function with a narrow transition region between the plasma and tissue phases.

\section{Results}
%
\subsection{Numerical verification for spherical diffusion}
\begin{figure}[htbp]
  \centering
    \hspace{1.5cm}\includegraphics[clip,width=16cm]{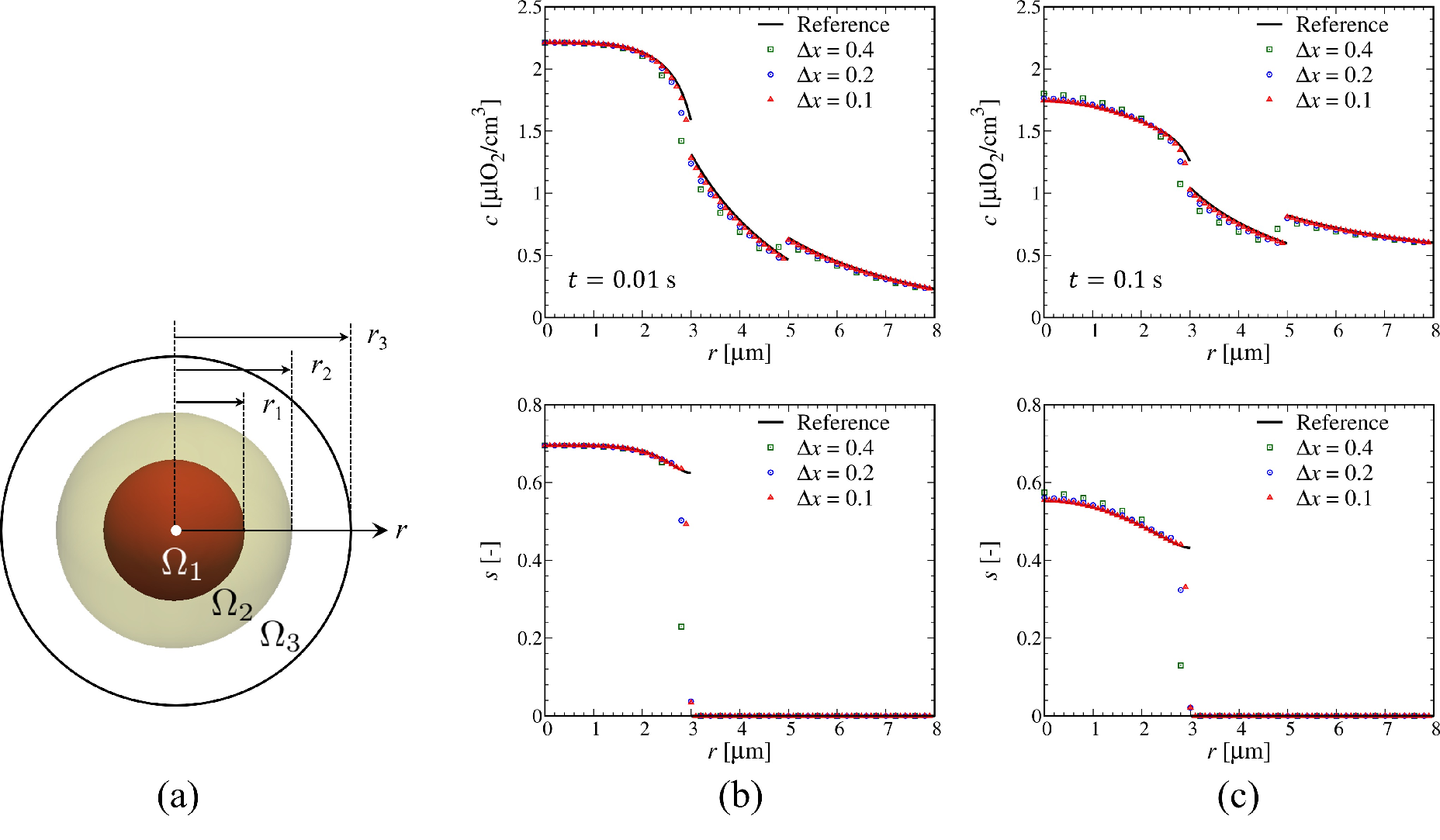}
    \caption{
    (a) Definition of the domains for a spherical diffusion problem: $\Omega_1$ is the internal RBC phase, $\Omega_2$ is the plasma phase, and $\Omega_3$ is the tissue phase.
    (b and c) Oxygen concentration $c$ as a function of the distance from the entrance of domain $r$ at $t = 0.01$ s and $0.1$ s, where solid black lines are the reference.
    The symbols are numerical results with various spatial resolutions $\Delta x$.
    }
    \label{fig:comparison}
\end{figure}

Assuming spherically symmetric diffusion without convection, the transport equations~\eqref{mix_c} and \eqref{mix_s} can be reduced to a 1D diffusion problem, where three computational phases $\Omega_1 \cup \Omega_2 \cup \Omega_3$ are defined with respect to the radial distance $r$, as shown in Fig.~\ref{fig:comparison}(a). Because the solution to such a spherically symmetric diffusion problem satisfies flux continuity along the outward normal direction at the interface, we use this solution to validate the accuracy of the discretized form of Eqs.~\eqref{mix_c} and \eqref{mix_s} in the full 3D problem. The spherically symmetric diffusion equations of $c$ and $s$ are written as
\begin{subequations}
\begin{eqnarray}
  && \frac{\partial c}{\partial t} = \frac{1}{r^2} \frac{\partial}{\partial r} \left( r^2 D \frac{\partial c}{\partial r} \right) + g, 
  \quad {\rm for} \ 0 \leq r \leq r_c, \label{c_sym} \\
  && \frac{\partial s}{\partial t} = \frac{1}{r^2} \frac{\partial}{\partial r} \left( r^2 D_s \frac{\partial s}{\partial r} \right) - f_{Hb}, 
  \quad {\rm for} \ 0 \leq r < r_1, \\
  && g = 
  \begin{cases}
  e f_{Hb} & \text{in} \ \Omega_1: \left\{ r| 0 \leq r < r_1 \right\}, \\
  0 & \text{in} \ \Omega_2: \left\{ r| r_1 \leq r < r_2 \right\}, \\
  -M_c & \text{in} \ \Omega_3: \left\{ r| r_2 \leq r \leq r_c \right\},
  \end{cases}
\end{eqnarray}
\end{subequations}
where $r_1$, $r_2$, and $r_c$ are the outer edge of the internal capsule phase, plasma phase, and computational domain, respectively. The symmetric and Neumann boundary conditions are used for the computational domain or interface, that is, $\partial_r c|_{r=0}=\partial_r c|_{r=r_c}=0$ and $\partial_r s|_{r=0}=\partial_r s|_{r=r_1}=0$. The interfacial conditions are also used between each phase. We consider that the internal region of a spherical capsule with radius $r_1$ = 3 $\mu$m is fixed at the center of a capillary (assumed to have spherical geometry) with $r_2$ = 5 $\mu$m. The capillary is embedded within a spherical tissue region with radius $r_3$ = 15 $\mu$m. The initial oxygen saturation is set to $s_0 = 0.7$ for $r \leq r_1$, and oxygen concentration $c_0 = 0$ for all domains $0 \leq r \leq r_c$. The reference solution is obtained from solving Eq.~\eqref{c_sym} at a fine spatial resolution $h_{ref}=\Delta r=0.015$ $\mu$m using a second-order finite difference method.

Full 3D simulations are performed in a cubic domain $x,y,z\in[-8,8]$ $\mu$m, where the Dirichlet boundary condition is imposed on the domain edge from the reference solution. The simulations are run for up to $0.1$ s at various spatial resolutions, $\Delta x = 0.4$, $0.2$, and $0.1$ $\mu$m. The time resolution $\Delta t$ is set to $\Delta t=5$ $\mu$s for $\Delta x=0.4$ and $0.2$, and $\Delta t=1.25$ $\mu$s for $\Delta x=0.1$ $\mu$m.

Figures~\ref{fig:comparison}(b) and \ref{fig:comparison}(c) show the comparison between the reference solution and the numerical results at two different time points: $t = 0.01$ s and $0.1$ s. Oxygen concentration diffuses from the capsule into plasma and eventually reaches the tissue phases as time progresses. The discontinuities on the interfaces on $r = r_1$ and $r_2$ arise from differences in solubility between the phases. Our numerical results demonstrate good agreement with the reference solution, particularly at the highest resolution $\Delta x = 0.1$ $\mu$m. Thus, the proposed mixture formulations accurately capture the distribution of oxygen concentration across phases characterized by different solubility and diffusivity. Moreover, the present solution accurately captures the profiles of oxygen saturation $s$: a smooth profile inside the capsule ($r < r_1$), interfacial jump at $r=r_1$, and  nearly zero constant profile outside the capsusle ($r>r_1$). Because the results obtained with $\Delta x = 0.2$ $\mu$m are comparable with those with the finest resolution $\Delta x = 0.1$ $\mu$m, we use $\Delta x = 0.2$ $\mu$m in the following analyses to avoid computational load. Hereafter, the time resolution is fixed: $\Delta t=5$ $\mu$s.

\subsection{Numerical example of a moving interface without membrane deformation} \label{sec:moving_interface}
We use a numerical experiment to investigate the effect of discretization for the advection term. To validate numerical results, we solve a diffusion problem in different frameworks of motion, with Lagrangian and Eulerian descriptions. A spherical capsule with radius $a=3$ $\mu$m is located in a straight tube with a diameter of $d=10$ $\mu$m in a cuboid domain $L_x=128$ $\mu$m, $L_y=32$ $\mu$m and $L_z=25.6$ $\mu$m. The straight tube is aligned parallel to the $x$ direction and the outside region is set to the tissue phase. If we consider a spatially uniform transport velocity for all phases ($\Omega_1$, $\Omega_2$, and $\Omega_3$), the result becomes the same as that without the velocity field. Although such cases are not distinguished in a physical sense, numerical treatments are very different with or without the velocity field because the interface is moving or fixed in the Eulerian mesh system used in this study.

\begin{figure}[htbp]
  \centering
    \includegraphics[clip,width=15cm]{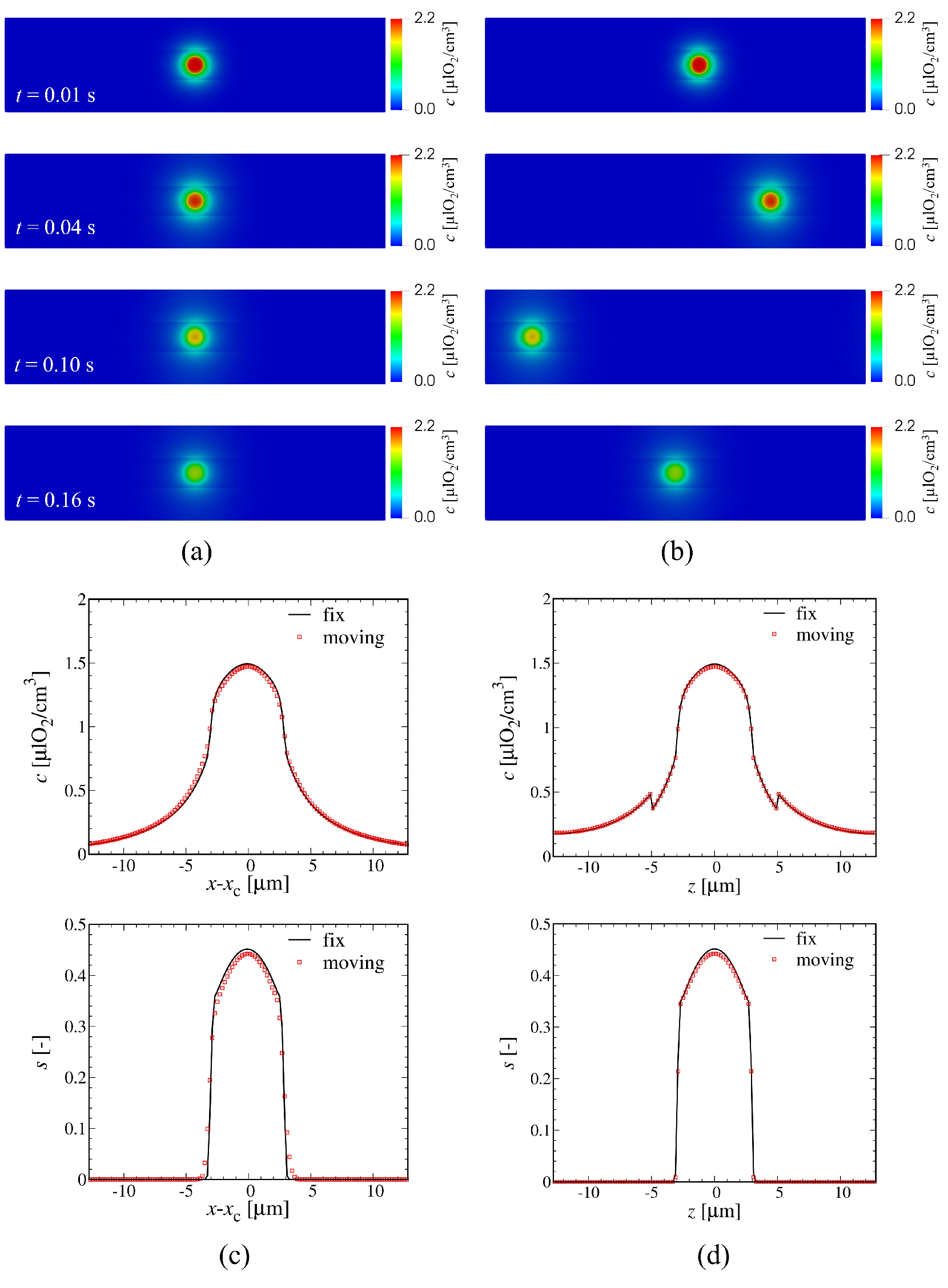}
    \caption{
    Numerical validation of oxygen transport with a moving (but non-deformable) interface.
    Instantaneous solutions of oxygen concentration $c$ for the fixed interface (a) and moving interface with a constant advection velocity (b).
    (c and d) Comparisons of axial profiles in both stream-wise $x$ and span-wise $z$ directions for $c$ and $s$ at $t=0.16$ s (after one period) for the fixed and moving interfaces.
    The results in panel (c) are shown as a function of the relative coordinate system based on the centroid of the spherical capsule, $x_c$.
    }
    \label{fig:validation_moving}
\end{figure}

Figures~\ref{fig:validation_moving}(a) and \ref{fig:validation_moving}(b) show snapshots of the numerical results for $c$ without the velocity (i.e., fixed interface) and with an $x$-wisely uniform velocity $U=0.8$ mm/s (i.e., moving interface), where the periodic boundary condition is imposed on the domain edges in the $x$ direction. The overall profiles are in good agreement with each other, and the oxygen transport into the tissue phase is well captured in the case with the moving interface. Figures~\ref{fig:validation_moving}(c) and \ref{fig:validation_moving}(d) compare axial profiles in both the stream-wise $x$ and span-wise $z$ directions for $c$ and $s$ at $t=0.16$ s (after one period) for the fixed and moving interfaces. The results in panel (c) are shown as a function of the relative coordinate system based on the centroid of the spherical capsule, $x_c$. The results for the moving interface capture the jumps for $c$ and $s$ across the phase interface; however the profile is slightly diffusive around the capsule interface in the $x$ direction (Fig.~\ref{fig:validation_moving}(c)) and a small amount of dissipation is observed inside the capsule phase. Despite this, the dissipation level of oxygen concentration $c$ is acceptable up to $t=0.16$ s, which is a reasonable time period to discuss phenomena in capillary networks.
In addition, it was demonstrated that, within the mixture formulation, evaluating the diffusion coefficient using the harmonic average yields superior accuracy compared to the arithmetic average (\ref{appendixA}).

\subsection{Oxygen transport with RBCs in a straight capillary}
An RBC is modeled as a biconcave capsule or a Newtonian fluid enclosed by a thin elastic membrane with a major diameter of 8 $\mu$m and maximum thickness of 2 $\mu$m. We define the initial shape of the RBC as a biconcave, where the initial (isotropic) stretch is set to $\lambda_s=1.05$ to avoid numerical instability caused by membrane wrinkles at the present spatial resolution $\Delta x=0.2$ $\mu$m. The RBC capsule is discretized using 2880 triangular elements. The usual distribution of the Hb concentration in individual RBCs ranges from 27 to 37 g dl$^{-1}$, corresponding to the cytoplasmic viscosity that is taken to be $5$--$15$ cP~\cite{Mohandas2008}, whereas the normal plasma viscosity is $\mu_{plasma} = 1.1$--$1.3$ cP for plasma at 37$^\circ$C. Thus, we set $\mu_1 = 6$ cP ($=6.0 \times 10^{-3}$ Pa$\cdot$s), which is five times higher than the plasma viscosity: $\mu_{plasma} = \mu_2 = 1.2 \times 10^{-3}$ Pa$\cdot$s~\cite{Harkness1970}.

\begin{figure}[htbp]
  \centering
    \includegraphics[clip,width=16cm]{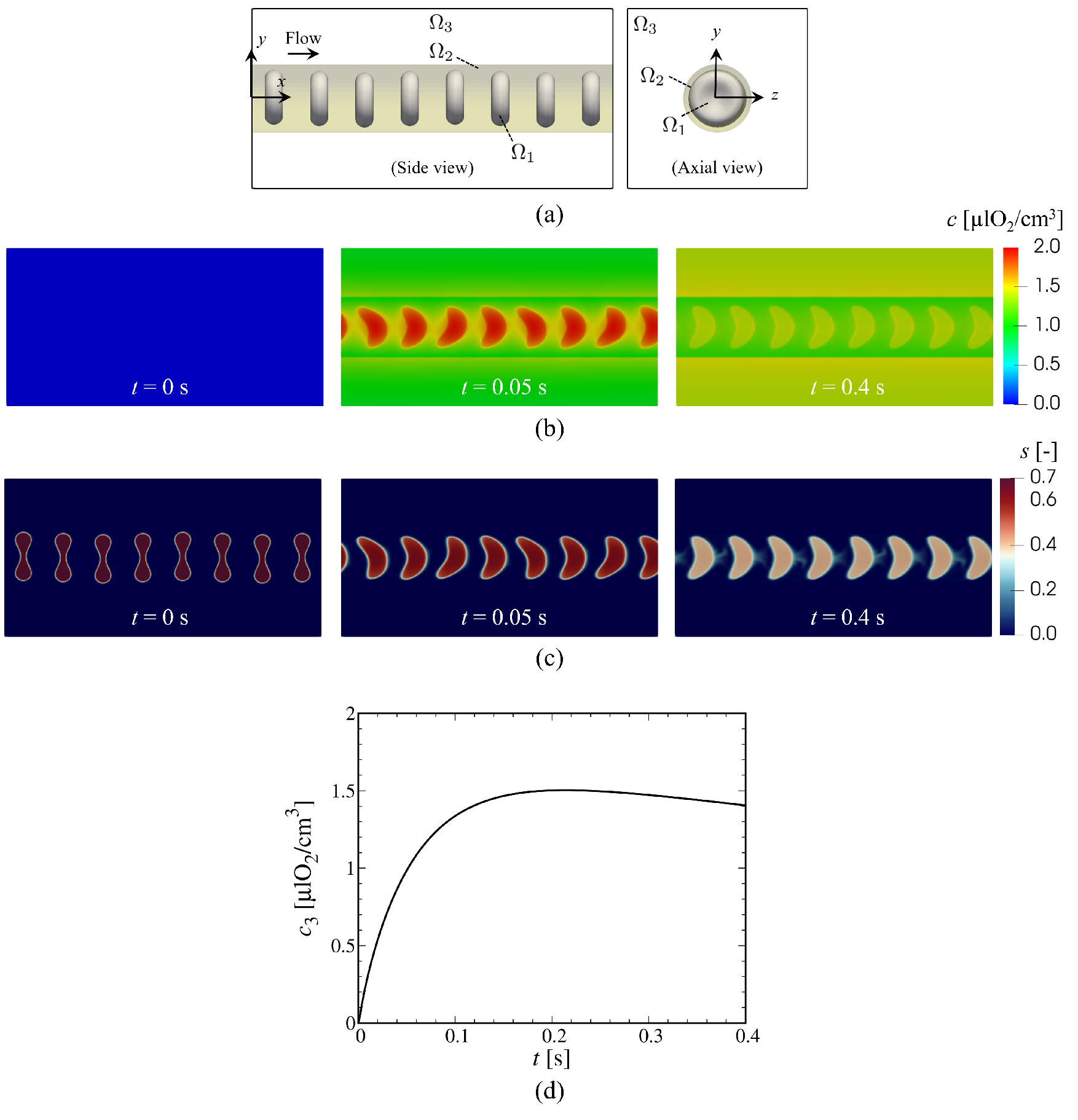}
    \caption{
    (a) Computational domains for oxygen transport with RBCs in a straight capillary, consisting of the internal RBC ($\Omega_1$), plasma ($\Omega_2$), and tissue ($\Omega_3$): ($left$) lateral view and ($right$) front view.
    The flow direction is from left to right.
    Snapshots of oxygen concentration $c$ (b) and saturation $s$ (c) on the lateral cross-sectional area ($z$ = 0) at ($left$) the initial state $t = 0$ s, ($middle$) $t = 0.05$ s, and ($right$) $t = 0.4$ s in $d = 9.66$ $\mu$m.
    (d) Time history of $c_3$ averaged in the tissue phase $\Omega_3$.
    }
    \label{fig:results_c_s}
\end{figure}

Now we investigate oxygen transport in multi-RBCs flow in a straight capillary, as shown in Fig.~\ref{fig:results_c_s}(a), where the capillary length $L$ is set to $L = 51.2$ $\mu$m and the volume fraction of RBCs, the so-called hematocrit $Hct$, is approximately $20$\%. The vessel diameter $d$ is set to be $d = 9.66$ $\mu$m, which is determined based on the $Hct$ and the number of RBCs $N_{RBC}$ ($\in \mathbb{Z}$), that is,
\begin{equation}
  d = \sqrt{\frac{4 N_{RBC} V_{RBC}}{\pi L}},
\end{equation}
where $V_{RBC}$ is the volume of an RBC ($\approx$ 96 $\mu$m$^3$ \cite{Evans1972}). The pressure gradient is set to $-dp/dx$ = 0.2 Pa/$\mu$m, and periodic boundary conditions are imposed on the flow direction. Therefore, in this problem, RBCs carrying HbO$_2$ periodically enter the capillary. The tissue region is set to be $25.6^2$ $\mu$m$^2$. The initial oxygen saturation is set to $s_0= 0.7$ in the internal RBC ($\Omega_1$). The initial oxygen concentration is zero ($c_0 =  0$) for the entire spaces ($\Omega_1 \cup \Omega_2 \cup \Omega_3$).

Figures~\ref{fig:results_c_s}(b) and \ref{fig:results_c_s}(c) show the distribution of oxygen concentration $c$ and saturation $s$ at $t=0.05$ s and $0.4$ s on the lateral cross-sectional area, respectively. The fully oxygenated Hb immediately diffuses through the RBC membrane, plasma and ultimately into tissues. Figure~\ref{fig:results_c_s}(d) shows the time history of $c_3$ averaged in the tissue phase $\Omega_3$. Corresponding to the decrease of $c_1$ (or $s_1$), $c_3$ steeply increases and reaches the maximum ($t \approx 0.2$ s), that is, diffusion and consumption have been balanced, then it decreases because the provided oxygen from RBCs has run out and consumption becomes the dominant event in tissue.

Numerically obtained the single-concave (or parachute-like) shape of RBCs shown in Figs.~\ref{fig:results_c_s}(b) and \ref{fig:results_c_s}(c) has been well investigated, e.g., in {\it in vivo} observations in capillaries with diameters of approximately 7 $\mu$m~\cite{Skalak1969}.
Numerical simulations have also successfully reproduced the shape, e.g., in multi-cellular interaction problems~\cite{Takeishi2014, Takeishi2015, Takeishi2017}, as well as the single cell level~\cite{Takeishi2021}.
These numerical studies further revealed the underling mechanism of the morphological transition after flow onset:
the RBC membrane near the tube wall experiences drug due to the no-slip boundary condition, whereas the membrane near the centerline flows faster,
resulting in a stable parachute shape.
The stable deformed shape potentially shifts to other morphologies depending on tube diameter, capillary numbers ($Ca$), and viscosity ratios, in association with the equilibrium position of the RBC~\cite{Takeishi2021}.

Oxygen saturation $s$ is seen to smear out around the front regions of RBCs at $t=0.4$ s (Fig.~\ref{fig:results_c_s}(c)).
This may be from numerical diffusions that result from solving the advection term. In a capillary tube, RBCs form a single-file line and cause bolus flow with vortex-like streamlines between RBCs~\cite{Takeishi2017} that results in high shear around the front interface of each RBC. As shown in our numerical example in the previous section, the distribution of $s$ (and $c$) across the interface is diffusive in the direction of advection (Fig.~\ref{fig:validation_moving}(c)). This diffusive distribution is artificially convected by the high shear flow. The fifth-order TENO scheme applied to the advection term is a less dissipative scheme; however, it cannot accurately capture both continuous and discontinuous profiles of $s$ and $c$ inside each material phase and their interface. 
Here, in the analysis shown in~Fig.~\ref{fig:results_c_s},
the mean velocity of RBCs is approximately $0.55$ mm/s, and thus, over a duration of $0.4$ s,
RBCs travel about $220$ $\mu$m along the capillaries (see also Fig.~\ref{fig:regions_3d_network}),
or is comparable to the total travel distance even when the lateral ($y$-direction) motion of RBCs is taken into account.
We therefore conclude that reasonable results can be obtained, at least, by time averaging over a period of $\leq 0.4$ s.

\subsection{Oxygen transport in simple capillary network models}
\begin{figure}[htbp]
  \centering
    \includegraphics[clip,width=13cm]{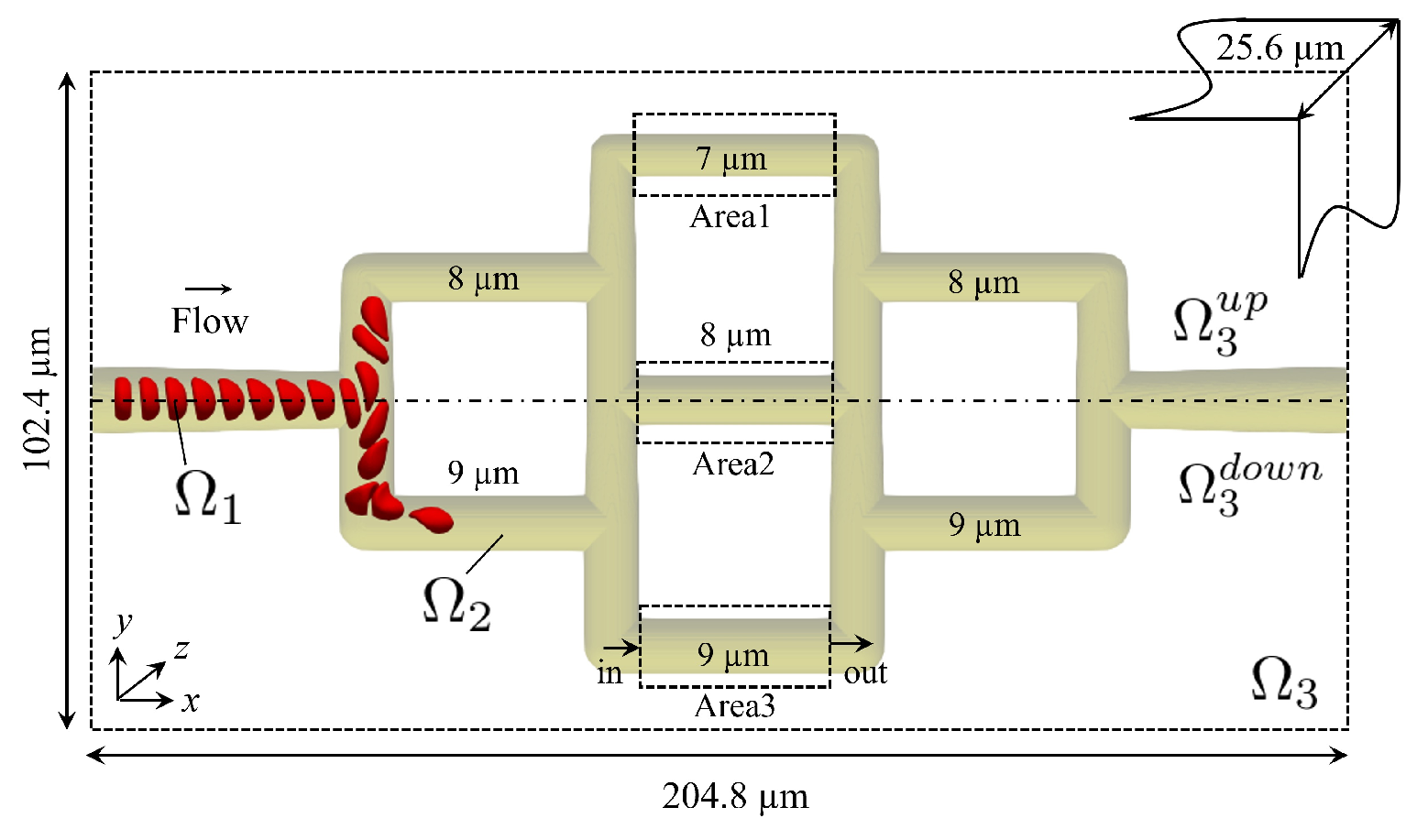}
    \caption{
    Microvascular network model, where the middle area consists of three capillaries of Area~1 for $d = 7$ $\mu$m, Area~2 for $d = 8$ $\mu$m, and Area~3 for $d = 9$ $\mu$m.
    The inlet vessel tapers off from $d = 11$ $\mu$m to $9$ $\mu$m, and the outlet vessel tapes out from $d = 9$ $\mu$m to $11$ $\mu$m.
    }
    \label{fig:regions_3d_network}
\end{figure}

To clarify the relationship between individual RBCs and tissue oxygenation in a microvascular system, we investigate oxygen transport in a simplified capillary network, as shown in Fig.~\ref{fig:regions_3d_network}. The computational domain is defined as a rectangular box of size $204.8$ $\mu$m$\times$ $102.4$ $\mu$m $\times$ $25.6$ $\mu$m along the stream-wise $x$, wall-normal $y$, and span-wise $z$ directions. The capillary network consists of a straight channel with diameters ranging from $d = 7$ $\mu$m to $9$ $\mu$m. A zero-flux (Neumann) condition ($\partial_n c = 0$) is imposed on the boundaries of the analysis domain $\Omega$. The initial oxygen concentration and HbO$_2$ saturation are set as follows: $c_0 = 0$ in all phases ($\Omega_1 \cup \Omega_2 \cup \Omega_3$) and $s_0$ = 0.7 in $\Omega_1$. The inserting (or adding) volume rate of RBCs at the inlet, $Q_a$ [$\mu$m$^3$/s], is determined by the adding hematocrit $Hct_a$ in the first vessel, defined as $Hct_a = Q_a/Q_{in}$, where $Q_{in} = U_{in} \pi d_{in}^2/4$ is the plasma flow rate at the inlet and $Q_a$ is calculated as $Q_a=V_{RBC}/T_a$, where $V_{RBC}$ [$\mu$m$^3$] is the RBC volume and $T_a$ [s] is the adding time of an RBC. We test four $Hct_a$ values: $0.112$ ($T_a=6$ ms), $0.168$ ($T_a=9$ ms), $0.224$ ($T_a=12$ ms), and $0.336$ ($T_a=18$ ms). $U_{in}$ is the inlet velocity assumed to be uniform in space and time, which is set to $U_{in}=5$ mm/s.

\begin{figure}[htbp]
  \centering
    \includegraphics[clip,width=17cm]{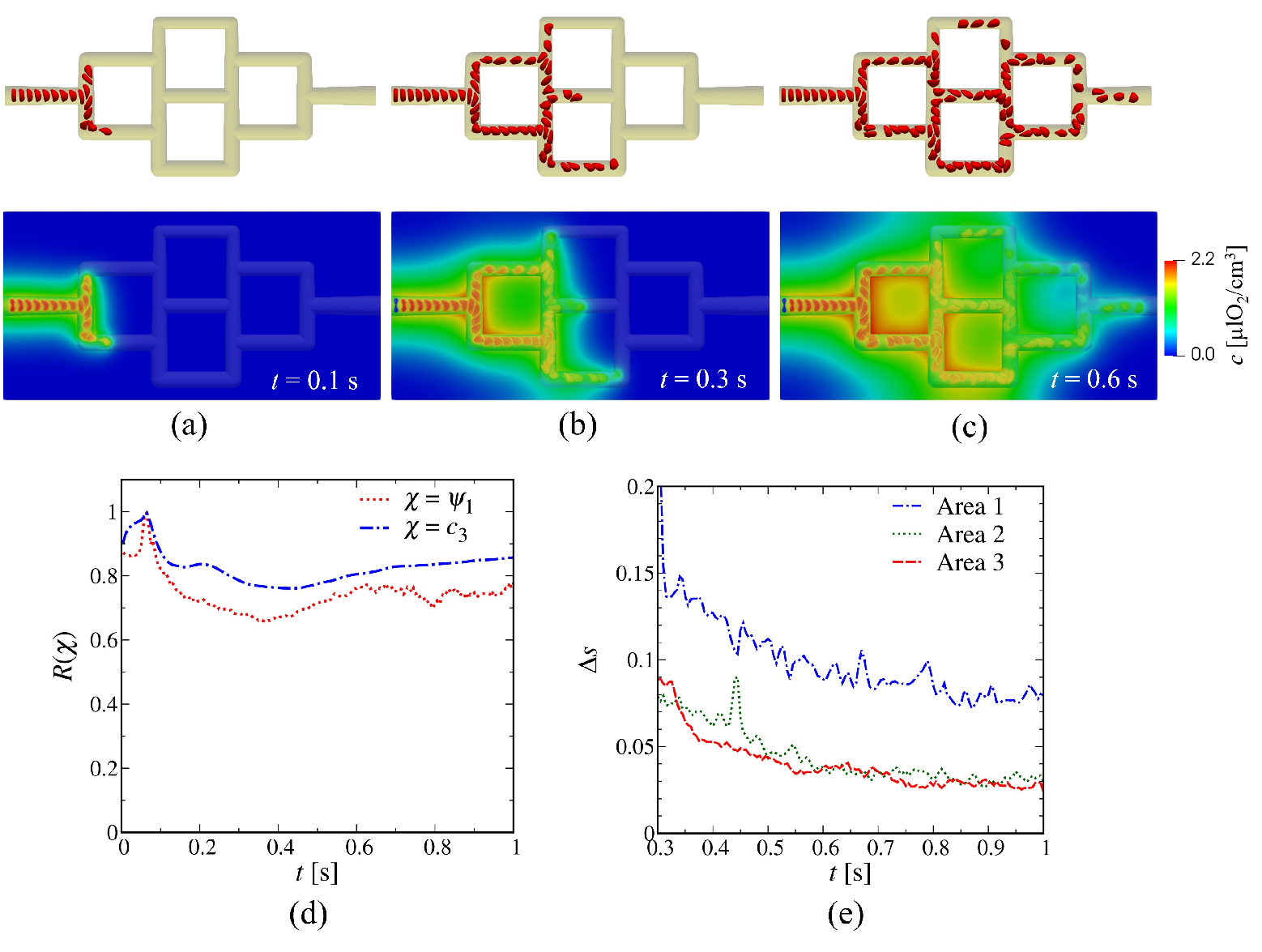}
    \caption{
    (a--c) Snapshots of the distribution of RBCs ($top$) and oxygen concentration $c_i$ ($bottom$) in $\Omega_1 \cup \Omega_2 \cup \Omega_3$ at $t = 0.1$ s (a), $0.3$ s (b), and $0.6$ s (c).
    (d) Time history of the ratio of oxygen concentration $R\langle c_3 \rangle$ and the RBC phase $R\langle \psi_1 \rangle$.
    (e) Time history of the difference in oxygen saturation $\Delta s$ for individual RBCs in Area~3 ($d = 9$ $\mu$m).
    The results were obtained with the largest $Hct_a$ ($= 0.336$) investigated in the study (also see also the supplementary video).
    }
    \label{fig:results_c_network}
\end{figure}

Figures~\ref{fig:results_c_network}(a)--(c) show snapshots of the RBC distributions within the network and the corresponding oxygen concentration fields at various time points for $Hct_a = 0.336$, the highest cell volume fraction considered in this study. RBCs initially travel through the largest capillary with $d = 9$ $\mu$m (Area~3); however, as the local pressure increases in Area~3, the flow pattern changes. Consequently, RBCs begin to flow into the smaller capillaries with $d = 7$ $\mu$m (Area~1) and $d = 8$ $\mu$m (Area~2) (also see the supplementary video). To quantify the relationship between the RBC distribution and the tissue oxygen concentration $c_3$, the centerline is defined in the computational domain, and the ratio of $c_3$ and $\psi_1$ between the phase regions above ($\Omega^{(up)}$) and below ($\Omega^{(down)}$) the centerline is calculated (Fig.~\ref{fig:regions_3d_network}):
\begin{align}
  & R\langle c_3 \rangle 
  = \frac{\text{min}\left(\langle c_3 \rangle_{\Omega_3^{(up)}}, \langle c_3 \rangle_{\Omega_3^{(down)}}\right)}{\text{max}\left(\langle c_3 \rangle_{\Omega_3^{(up)}}, \langle c_3 \rangle_{\Omega_3^{(down)}}\right)}, \label{R_c3} \\
  & R\langle \psi_1 \rangle
  = \frac{\text{min}\left(\langle \psi_1 \rangle_{\Omega^{(up)}}, \langle \psi_1 \rangle_{\Omega^{(down)}}\right)}{\text{max}\left(\langle \psi_1 \rangle_{\Omega^{(up)}}, \langle \psi_1 \rangle_{\Omega^{(down)}}\right)}, \label{R_psi1}
\end{align}
where $\langle \cdot \rangle_{\Omega^{(up/down)}}$ is the spatially averaged quantity in $\Omega^{(up)}$ or $\Omega^{(down)}$. Hence, a completely homogeneous distribution is represented by $R \langle \cdot \rangle = 1$. Figure~\ref{fig:results_c_network}(d) shows the time histories of $R \langle c_3 \rangle$ and $R \langle \psi_1 \rangle$, which evaluate the homogeneity of tissue oxygenation and RBC distribution, respectively. The result shows that $c_3$ decreases until $t=0.3\sim0.4$ s and then increases over time, which suggests that the RBC distribution and oxygen supply/consumption in the tissue become balanced in each region.

The contribution of individual RBCs to the oxygen supply is quantified by the oxygen saturation difference in each capillary segment (Area~$k(=1,2,3$) in Fig.~\ref{fig:regions_3d_network}), $\Delta s^{(k)}$, given as
\begin{equation}
  \Delta s^{(k)}  = s^{(k)}_{in} - s^{(k)}_{out}
  \approx {\Delta t}^{(k)}_{pass} \langle{f}\rangle_{\Omega_1^{(k)}}
  = \frac{L^{(k)} \langle{f}\rangle_{\Omega_1^{(k)}}}{\langle{v_{RBC}}\rangle_{\Omega_1^{(k)}}},
  \ (k=1,2,3),
\end{equation}
where $s^{(k)}_{in}$ and $s^{(k)}_{out}$ represent the oxygen saturation levels at the inlet and outlet of Area~$k$, and $\langle{f}\rangle_{\Omega_1^{(k)}}$ and $\langle{v_{RBC}}\rangle_{\Omega_1^{(k)}}$ are the volume averaged deoxygenation rate and RBC velocity, respectively, in $\Omega_1$ for Area~$k$. ${\Delta t}^{(k)}_{pass}$ is the time interval during which the RBC passes through each Area~$k$, which is evaluated by dividing the axial distance of Area~$k$, $L^{(k)}$, by the average RBC velocity. A large $\Delta s$ indicates higher oxygen extraction from individual RBCs. Figure~\ref{fig:results_c_network}(e) shows the time history of $\Delta s$ in the largest capillary with $d$ = 9 $\mu$m (Area~3).The results demonstrate that RBCs passing through this capillary at earlier time points ($t \lessapprox 0.4\sim0.6$ s) release much more oxygen than those passing through later, and the individual $\Delta s^{(k)}$ values tend to saturate over time.

\begin{figure}[htbp]
  \centering
    \includegraphics[clip,width=17cm]{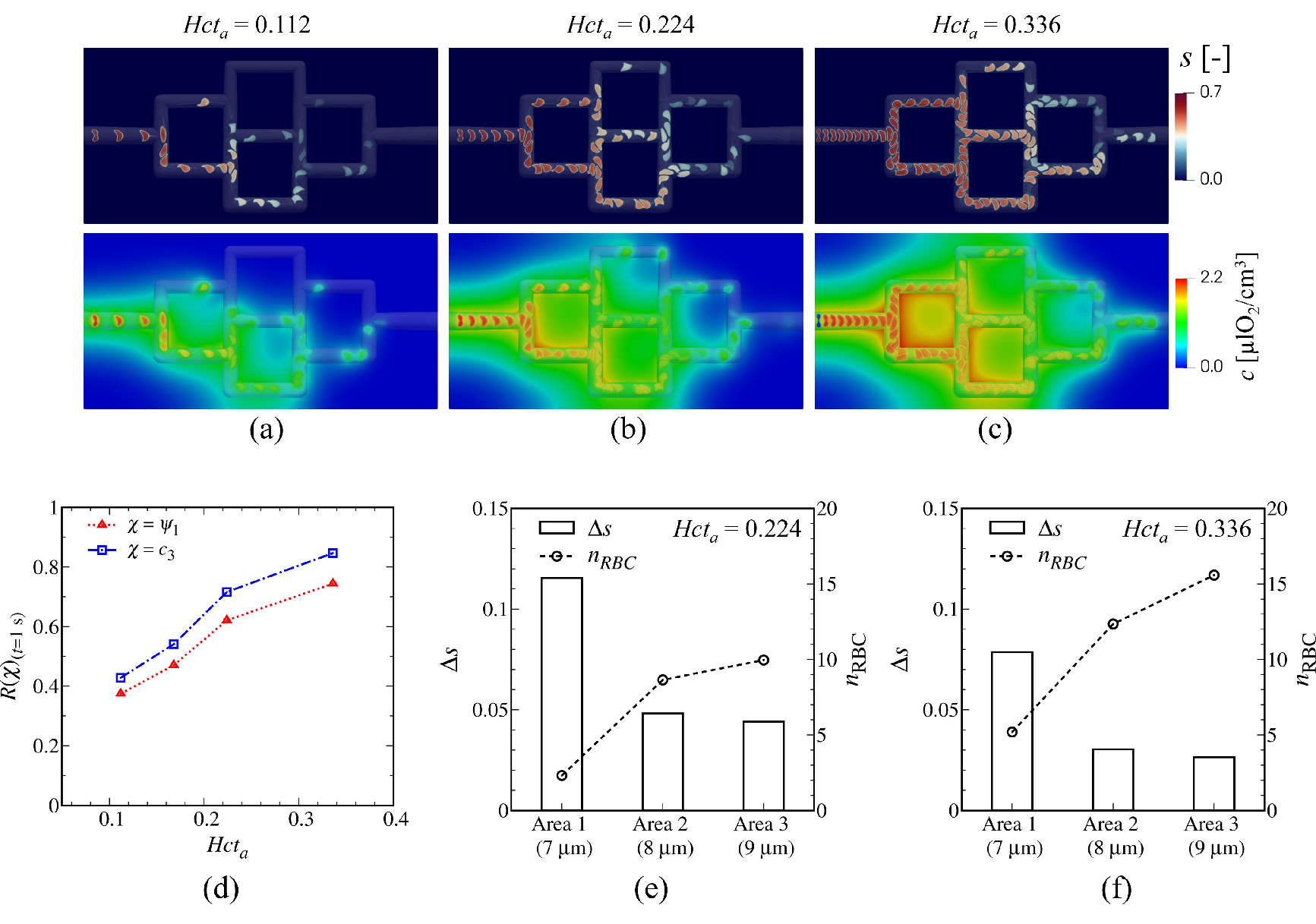}
    \caption{
    (a--c) Snapshots of oxygen saturation $s$ and oxygen concentration $c_i$ at $t = 0.6$ s for $Hct_a = 0.112$ (a), $Hct_a = 0.224$ (b), and $Hct_a = 0.336$ (c).
    (d) Ratio of oxygen concentration $R \langle c_3 \rangle$ and RBC phase $R \langle \psi_1 \rangle$ averaged over $0.8-1.0$ s.
    (e and f) Time-averaged oxygen extraction rates $\Delta s$ and the number of RBCs ($n_{RBC}$) passing through each Area for $Hct_a=0.244$ and $0.336$, where time averaging is performed over $0.2$ s between $t = 0.8$ and $1$ s.
    }
    \label{fig:results_c_network_HctD}
\end{figure}

Because homogeneous tissue oxygenation requires a sufficient number of RBCs, the value of $R \langle c_3 \rangle$ is expected to depend on $Hct_a$. Therefore, we performed simulations for various $Hct_a$. Figures~\ref{fig:results_c_network_HctD}(a)--\ref{fig:results_c_network_HctD}(c) show the oxygen concentration distributions for different $Hct_a$ under fully developed flow conditions ($t \approx 0.6$ s), defined as the time point when RBCs initially entering the domain have reached the outlet at the highest $Hct_a$ ($= 0.336$).
At lower $Hct_a$ ($= 0.112$), RBCs do not flow into the smallest capillary with $d = 7$ $\mu$m (Area~1), whereas at $Hct_a = 0.224$, the flow preference shifts and RBCs begin to perfuse smaller capillaries. These results suggest that effective RBC-mediated flow regulation within microvascular networks requires a minimum $Hct_a$ threshold.

The effect of $Hct_a$ on tissue oxygenation is quantified using the $c_3$ ratio~\eqref{R_c3} for each $Hct_a$, as shown in Fig.~\ref{fig:results_c_network_HctD}(d). In the same panel, the ratio of RBC distribution $R\langle{\psi_1}\rangle$ is superposed for comparison. As $Hct_a$ increases, tissue oxygenation becomes more homogeneous, which is indicated by $R\langle{c_3}\rangle$ approaching $1$. Because $R\langle{c_3}\rangle$ follows the same trend as $R\langle{\psi_1}\rangle$, tissue oxygenation can be estimated well from the local $Hct$ distribution. Given the strong correlation between the cell distribution (i.e., local $Hct$) and tissue oxygenation, the capillary diameter can provide a good approximation of the RBC flow rate. This finding supports the validity of continuum models in microvascular networks, such as that presented in~\cite{Hartung2018}. However, knowledge alone is insufficient to fully explain how RBCs can continue delivering oxygen downstream in the capillary network without depleting their HbO$_2$ content. Therefore, the contribution of individual RBCs to tissue oxygenation is quantified by the oxygen extraction rate, represented by the ensemble average of $\Delta s$ for $t\in[0.8,1]$ s in the quasi-steady state. The number of RBCs that pass through each Area~$k$ is also calculated by
\begin{equation}
  n_{RBC} = \frac{Q_{RBC} \Delta t_{interval}}{V_{RBC}},
\end{equation}
where $\Delta t_{interval}=0.2$ s is the time interval for the evaluation period from $0.8$ to $1$ s. Figures~\ref{fig:results_c_network_HctD}(e) and \ref{fig:results_c_network_HctD}(f) show both the oxygen extraction rate $\Delta s$ and the number of RBCs $n_{RBC}$ passing through each Area~$k$ at $Hct_a=0.224$ and $0.336$. As expected, the value of $n_{RBC}$ increases with $Hct_a$. However, $\Delta s$ tends to increase as $n_{RBC}$ decreases, which suggests that RBCs can autonomously regulate the oxygen supply to tissues in response to the local tissue oxygenation level.

\begin{figure}[htbp]
  \centering
    \includegraphics[clip,width=17cm]{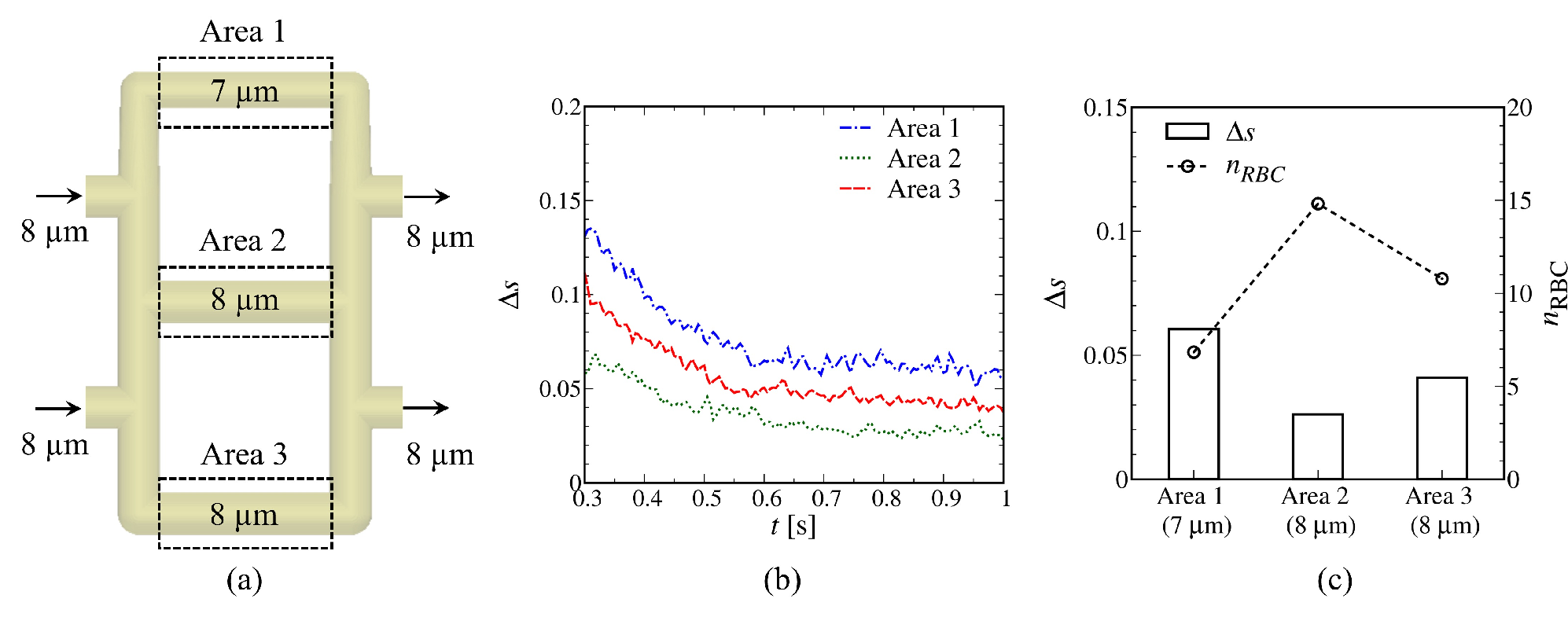}
    \caption{
    (a) Snapshot of the analysis area in the modified microvascular network, which consists of one small vessel (Area~1: $d = 7$ $\mu$m) and slightly larger vessels (Areas~2 and 3: $d = 8$ $\mu$m).
    (b) Time history of the difference in oxygen saturation $\Delta s$ for individual RBCs passing through each Area.
    (c) Time-averaged oxygen extraction rate $\Delta s$ and the number of RBCs passing through $n_{RBC}$ in each Area at $Hct = 0.336$, where time averaging is performed over $0.2$ s between $t= 0.8$ and $1$ s.
    }
    \label{fig:results_s_network}
\end{figure}

We perform a similar analysis with a different capillary network consisting of one small vessel with $d = 7$ $\mu$m and two slightly larger vessels with $d = 8$ $\mu$m (Fig.~\ref{fig:results_s_network}(a)). Figure~\ref{fig:results_s_network}(b) shows the time history of the oxygen saturation difference $\Delta s$ in each capillary for Area~$k$ ($k=1,2,3$), and Fig.~\ref{fig:results_s_network}(c) shows $\Delta s$ and $n_{RBC}$ for each Area~$k$ at $Hct_a = 0.336$. Compared with the previous network, a similar tendency is observed that $\Delta s$ in the smallest capillary ($d = 7$ $\mu$m) is constantly larger than in relatively larger capillaries ($d = 8$ $\mu$m), whereas $n_{RBC}$ is smaller than that in the other two Areas. However, the relationship between $n_{RBC}$ and $\Delta s$ among capillary segments (for each Area) alters when the capillary diameter in the bottom domain shrinks slightly (from $d=9$ to $8$ $\mu$m). The geometrical homogeneity of the capillary network enhances the homogeneity of the RBC distribution, followed by homogeneous tissue oxygenation (also see the supplementary video). These results suggest that RBCs can adaptively regulate the oxygen supply to tissues based on the surrounding oxygen tension $P_{O2}$.

\begin{figure}[htbp]
  \centering
    \includegraphics[clip,width=17cm]{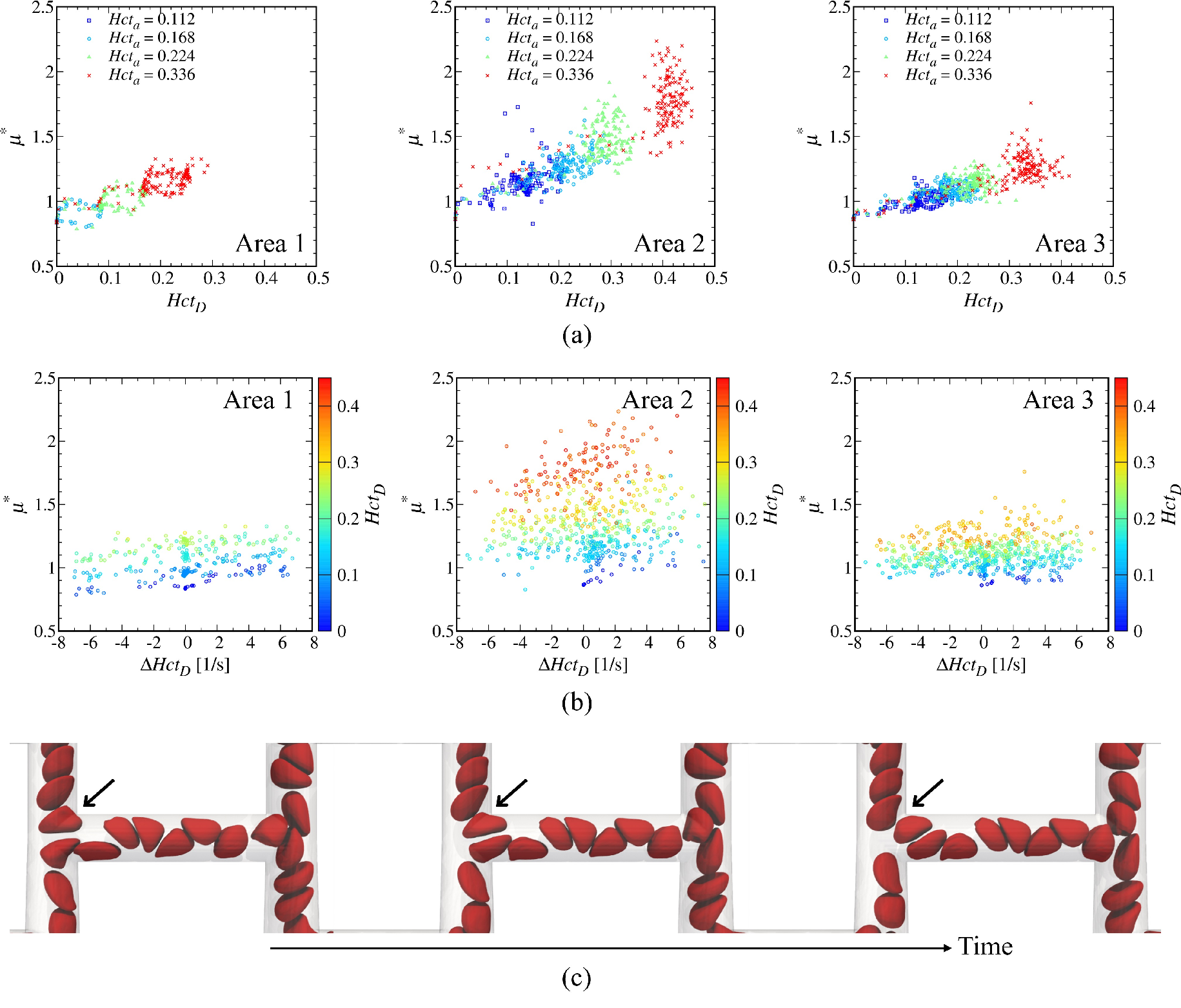}
    \caption{
    (a and b) Relationships between the apparent viscosity $\mu^{\ast}$ and the discharged hematocrit $Hct_D$ (a), and time derivative of $Hct_D$, $\Delta{Hct_D}$ (b), in each capillary segment for Areas~1--3.
    (c) Instantaneous solutions of the RBCs entering the capillary segment of Area~2, where the RBCs are highly deformed and interact with each other at the entrance of the segment.
    }
    \label{fig:results_viscosity}
\end{figure}

Figure~\ref{fig:results_viscosity}(a) shows the relationships between the discharged hematocrit $Hct_D$ and apparent viscosity $\mu^{\ast}$ on each capillary segment in Area~$k$ ($k=1,2,3$) defined as
\begin{align}
  & Hct_D = \frac{Q_{RBC}}{Q}, \\
  & \mu^{\ast} = \frac{\pi d^4 \Delta{p}}{128 L Q}\frac{1}{\mu_{plasma}},
\end{align}
where $Q$ is the volumetric flow rate of fluid, $\Delta p$ is the pressure difference between the inlet and outlet, and $L$ is the axial length, with regard to each Area. Data for $t\in[0.5,1]$ s are used. Positive correlations are observed between $Hct_D$ and $\mu^{\ast}$ for all Areas, and different RBC-inserting conditions $Hct_a$ are well organized by each single correlation curve in each Area. However, there are variations in $\mu^{\ast}$ at the same $Hct_D$, which indicates that other factors inhere to determine $\mu^{\ast}$; that is, flow resistance is not only determined by the steady configuration of RBCs passing through the capillary segment. Figure~\ref{fig:results_viscosity}(b) shows the relationship between the time derivative of $Hct_D$ as $\Delta{Hct_D} \equiv d{Hct_D}/dt$ and $\mu^{\ast}$. Good correlations are observed between them, which shows that the increase of the temporal change of $Hct_D$ increases the apparent viscosity $\mu^{\ast}$ (i.e., flow resistance). When $\mu^{\ast}$ is approximated with a function as $\mu^{\ast}_h = F(Hct_D, \Delta{Hct_D})=a\cdot{Hct_D}+b\cdot{\Delta{Hct_D}}+c$, the coefficient of determination $R^2$ is calculated with fitting as $R^2_{{\rm Area} 1}=0.646$, $R^2_{{\rm Area} 2}=0.793$, and $R^2_{{\rm Area} 3}=0.896$ for each Area, thereby demonstrating good correlations. These results indicate that dynamic effects are important for determining the local flow resistance in capillary networks, which is not seen in fully developed flows in straight tubes~\cite{Pries1992}. The fully developed relation between $Hct_D$ and $\mu^{\ast}$ is known as the Fahraeus--Lindqvist effect and widely applied to the modeling of the local change of tube resistance in existing microcirculation models (e.g.,~\cite{Koppl2020}). The present results rebuild the empirical relation by adding dynamic effects. Correlation behaviors among the Areas are not common where the gradient in Area~2 is larger than Areas~1 and 3. This could be caused by the fact that RBC-entering behaviors into the capillary segment in Area~2 are different from those in other cases (Area~1 and 3) because the RBCs come from two vertical sides and the scenario is slightly different from that in the other Areas. In Area~2, many RBC interactions coming from both sides frequently occur, which results in high flow resistance (Figure~\ref{fig:results_viscosity}(c)). This may cause the large variations in $\mu^{\ast}$ for Area~2.

\subsection{Effect of the existence of the membrane on tissue oxygenation}
The question here is how the RBC membrane contributes to oxygen transport, that is, why the membrane is essential. Insights into this question are particularly valuable for the design of the artificial oxygen carrier system. Most previous mathematical analyses of oxygen transport have neglected resistance in the radial direction and have modeled blood as a continuum with Hb homogeneously distributed in plasma, rather than being localized within RBCs. The importance of the RBC membrane in oxygen transport was addressed in pioneering experimental work by Hartridge and Roughton~\cite{Hartridge1927}, who showed that a specific oxygenation process took 16 times longer in whole blood than in a solution of Hb (i.e., Hb not encapsulated within RBCs)~\cite{Hartridge1927}. This indicates that the rates of oxygen transport in blood may be significantly lower than in an Hb solution with the same Hb content. This finding was later supported by a theoretical study by Hellums~\cite{Hellums1977}, who showed that $P_{O_2}$ gradients in capillaries with diameters similar to RBCs resulted in substantially lower oxygen transport rates in blood than a homogenous solution with an equivalent Hb content~\cite{Hellums1977}. Although these studies confirmed the importance of the RBC membrane in oxygen transport, to date no research has clearly identified how the membrane or the pattern of HbO$_2$ inflow affects tissue oxygenation in dynamic conditions. Thus, we address this issue using our developed diffuse interface approach.

\begin{figure}[htbp]
  \centering
    \includegraphics[clip,width=17cm]{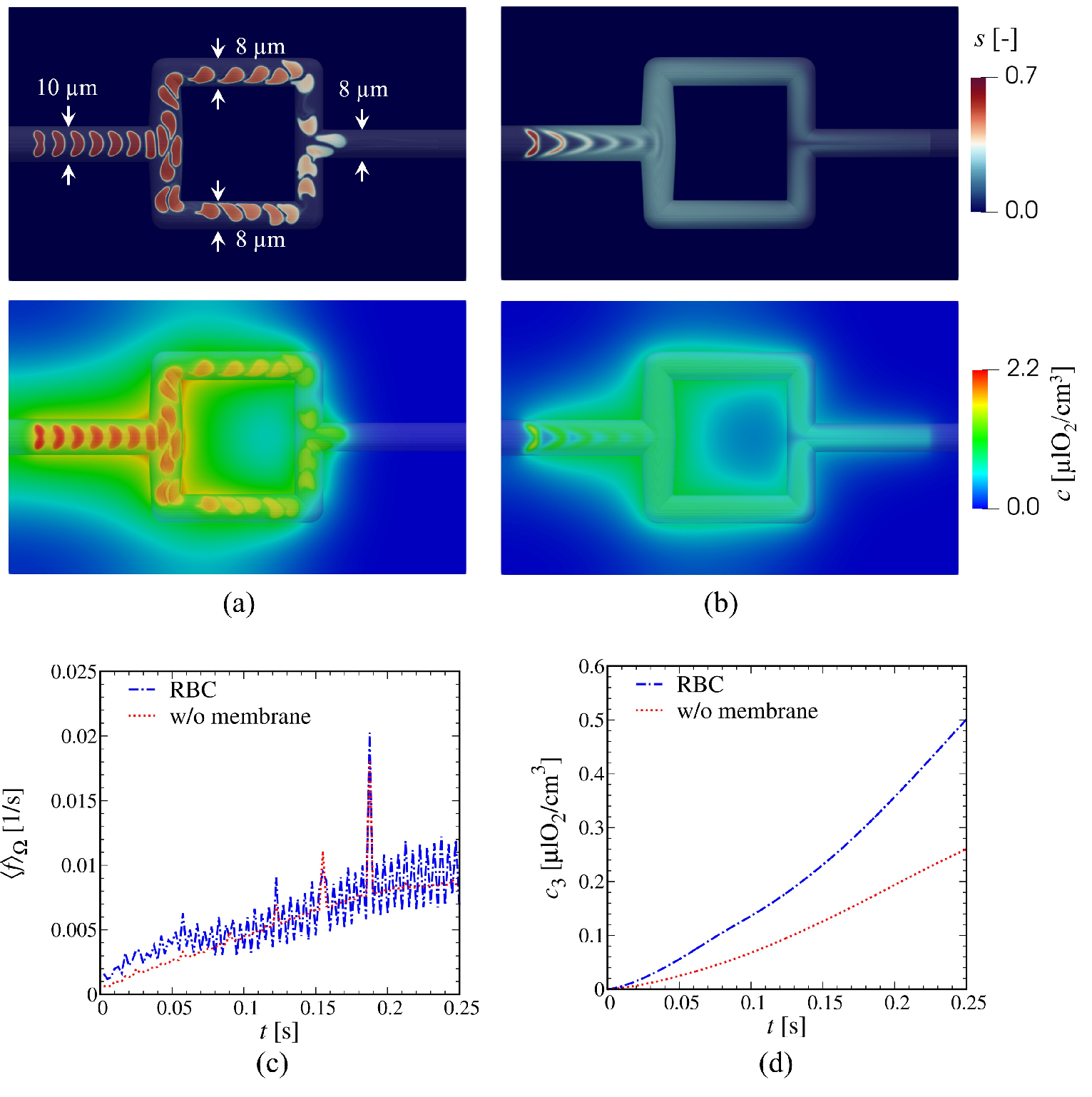}
    \caption{
    (a and b) Snapshots of the distribution of oxygen saturation $s$ and oxygen concentration $c$ at $t = 0.25$ s for various analysis cases in the simple microvascular network.
    The diameter of the inlet straight vessel is $10$ $\mu$m and others are 9 $\mu$m.
    Two different cases are considered: (a) existence of the RBC and (b) absence of the RBC membrane.
    (c) Time histories of the oxygen release rate $\langle{f}\rangle$ and (d) oxygen concentration in tissue $c_3$ for both cases.
    }
    \label{fig:simple_network}
\end{figure}

For simplicity, we calculate the transport of HbO$_2$ without considering membrane dynamics, that is, Hb transport including the advection and diffusion occurs in the plasma phase of an idealized microvascular network, as shown in Figs.~\ref{fig:simple_network}(a) and \ref{fig:simple_network}(b). The network consists of one bifurcation and one confluence within a rectangular box of size $128$ $\mu$m $\times$ $76.8$ $\mu$m $\times$ $25.6$ $\mu$m along the stream-wise $x$, wall-normal $y$, and span-wise $z$ directions. The diameters of the inlet and outlet straight capillary are $d = 10$ $\mu$m while the other vessels have diameters of $d = 8$ $\mu$m (Figs.~\ref{fig:simple_network}(a) and (b)). The initial conditions, boundary treatment, and numerical parameters are identical to those described in the previous subsection. For the case without the RBC membrane, we introduce the Hb-dissolved plasma phase as $\Omega_1\cup\Omega_2$ with no distinction (or jump conditions) between the RBC phase and plasma phase, and the transport of $s$ is calculated by Eq.~\eqref{mix_s} in the Hb-dissolved plasma phase, where the jump conditions for $c$ and $s$ are imposed on the interface between the Hb-dissolved plasma and tissue with the following parameters: $\mu_1=\mu_2=\mu_{plasma}$, $\alpha_1=\alpha_2$, $D_1=D_2$, $\beta_1=\beta_2=1$, and $D_{s1}=D_{s2}=D_{Hb}$. In this study, the RBC or oxygen saturation in the region of RBC ($s_0=0.7$) is periodically added following $Hct_a = 0.336$.

Figures~\ref{fig:simple_network}(a) and \ref{fig:simple_network}(b) show snapshots of the spatial distribution of oxygen concentration $c$ and oxygen saturation $s$ for the two cases at $t = 0.25$ s. The results show that in both cases, $s$ gradually decreases from the upstream to the downstream regions, where the case without the membrane has more diffusive profiles. Because the rate of oxygen release from HbO$_2$ depends on the reaction rate $f$ (as defined in Eq.~\eqref{f}), the spatial average of the reaction rate $\langle{f}\rangle$ reflects the potential growth rate of the oxygen concentration in Hb-dissolved plasma. Figure~\ref{fig:simple_network}(c) shows the time history of $\langle{f}\rangle$ for two cases, where both levels are not significantly different. By contrast, Fig.~\ref{fig:simple_network}(d) shows that tissue oxygenation, represented by $c_3$, is higher in the case with the RBC membrane than in that without the membrane, even if the level of the oxygen release rate $\langle{f}\rangle$ is similar. It is expected that the diffusion of oxygen molecules into tissue would depend on the radial gradient of the oxygen concentration (i.e., diffusion flux). Our results clearly that show the case with the RBC membrane forms the cell-depleted peripheral layer and radial gradient of the oxygen concentration. We conclude that the RBC membrane plays a key role in preserving high oxygen concentration inside the RBC to maintain the high diffusion flux, thereby aiding efficient oxygen delivery to surrounding tissue.

\subsection{Limitations and future perspectives}
Although the proposed mixture formulation can handle transport problems for largely moving and deforming interfaces of flexible capsules, numerical dissipation occurs because of the process of solving the advection problem. This would be improved if state-of-the-art schemes were used to suppress numerical diffusion. A series of boundary variation diminishing methods~\cite{Sun2016, Wakimura2024} is a possible approach, which can switch reconstruction functions to capture both the continuous and discontinuous profiles in the evaluation of numerical fluxes. This would dramatically improve the numerical accuracy for the transport problem of the oxygen saturation, which is smooth in the RBC, but discontinuous across the interface. This modification would allow us to extend the proposed model to analyses using the real-world microvascular geometries, in which longer-range transport in both space and time could be considered. 

We considered a simple capillary network; however, in vivo, microcirculations form complex networks that can affect tissue oxygenation and are closely related to organ-specific functions~\cite{Augustin2017}. Researchers have shown that interactions among microvascular systems reduce $P_{O_2}$ heterogeneity when RBC velocity and inlet oxygen concentration vary across individual capillaries~\cite{Popel1989, Li2019}. Therefore, it is interesting to investigate how RBC transit time in physiologically relevant microvascular networks influences tissue oxygenation under varying capillary densities. Recent experimental studies have explained the relationship between RBC deformation and the export of vasoactive mediators (e.g., adenosine triphosphate), in addition to the mechanistic link between nanoscale oxygen unloading from Hb and the macroscale membrane deformation~\cite{McMahon2019}. However, the present numerical model does not incorporate the feedback mechanisms initiated by membrane deformation that influences oxygen release. Thus, a future challenge will be to model the mechanistic link between RBC deformation and oxygen export.

The proposed mixture formulations can be extended to various nutrients or the substance transport problem~\cite{Buerk2001}, including drug delivery via nanocarriers. As such, the developed methodology provides a framework for the investigation of the metabolic state of tumors, which are usually hypoxic and nutrient-deprived because of their vessels' malfunction~\cite{Carmeliet2011}.
From a medical therapeutic perspective,
lipid-coated microbubbles have recently attracted significant attention and can be employed in various therapeutic applications,
such as ultrasonic contrast agents for rapid imaging~\cite{Abu-Nab2025}.
Although further technical improvements are required, particularly for large density contrasts,
the present numerical model provide basis for further studies on dynamics of liquid-coated microbubles subjected to ultrasound in the blood flow.
When the pressure decreases from a high level,
there is a pathological risk of vaporization of dissolved nitrogen within the plasma, the so-called decompression sickness~\cite{Abu-Nab2025},
particularly among compressed-air workers, pilots divers, and astronauts.
Solving the problem of bubble expansion and development in complex pipelines can provide insights into this pathology,
which is however future study.

\section{Conclusions}
We presented a computational approach to model oxygen transport involving chemical reactions, convection, diffusion and metabolic consumption in tissues under cellular flow dynamics. Our approach is based on a diffuse interface method, which allows the seamless calculation of oxygen concentration in the cytoplasm (internal fluid) of RBCs, plasma (external fluid), and tissue regions. Oxygen transport is formulated as an advection-diffusion-reaction equation, and all governing equations are rewritten in mixture forms using indicator functions. These functions are introduced to distinguish each phase (plasma, cytoplasm, and tissue region) and to avoid the complexity of discontinuities across interfaces. The proposed approach accurately captures the analytical solution for spherically symmetric diffusion, and successfully demonstrates oxygen transport in both straight capillaries and their networks. Therefore, our numerical model offers a novel technique for elucidating how individual cell behavior relates to oxygen metabolism in microvascular systems.

The proposed mixture formulation is not limited to the non-conforming mesh system; it can also be applied to confirming (boundary-fitted, unstructured) mesh systems. Compared with the sharp interface approach, the diffuse interface does not require explicit coupling between quantities in different phases through interface conditions because the conditions are already incorporated into the mixture equations. This results in superior features for achieving numerical stability and efficiency when in solving the entire system compared with the sharp interface approach. It is expected that the boundary-fitted mesh will increase the local spatial resolution with aligned mesh cells around the interface and the numerical solution will be more accurate, even when using a diffuse interface.

The proposed method can be extended to gain insights not only into mechanisms underlying organ- or region-specific functions but also into designing artificial oxygen or drug carrier systems. Moreover, the method is not restricted to biophysical problems; it can be generalized to mass transfer problems in heterogeneous media.

\section*{Acknowledgments}
N.T. acknowledges funding from the Japan Society for the Promotion of Science (JSPS) KAKENHI JP25K07570 and JKA promotion funds from AUTORACE (2025M-480). Parts of simulations were performed using the computational resources of the supercomputer Fugaku, provided by the RIKEN Center for Computational Science (project ID: hp240080, hp240458) and the supercomputer ``Flow'' at the Information Technology Center, Nagoya University. The authors thank Shu Takagi (The University of Tokyo), Kazuyasu Sugiyama (The University of Osaka), Huanxiong Huang (York University), and Xiabo Gong (Shanghai Jiaotong University) for fruitful discussions about mixture formulations, and Kazuto Masamoto (The University of Electro-Communications) for constructive comments on RBC distributions in cerebral microcirculation. We thank Edanz (https://jp.edanz.com/ac) for editing a draft of this manuscript.

\clearpage

\section*{Figures}

\clearpage

\begin{appendix}
\setcounter{figure}{0}

\section{} \label{appendixA}
\begin{figure}[htbp]
  \centering
    \includegraphics[clip,width=17cm]{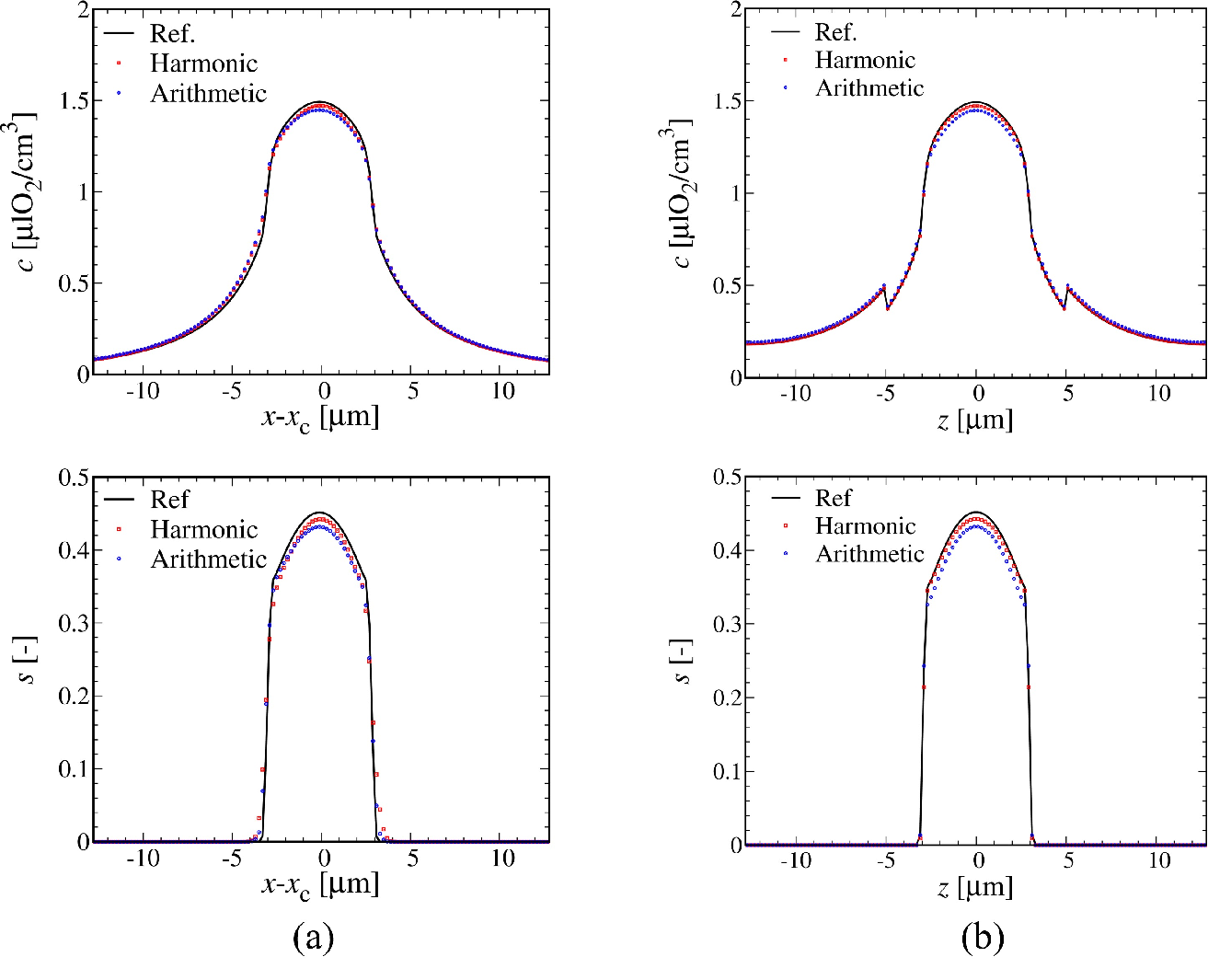}
    \caption{
    Comparison between approximations of the diffusion coefficient in the mixture formulation based on the harmonic average $D^{h}$ in \eqref{Dh} and the arithmetic average $D$ is presented. Results are evaluated for an oxygen-transport problem involving a non-deformable moving interface, as described in Section~\ref{sec:moving_interface}, and compared with those obtained using the fixed-interface condition as the reference (Ref). Axial profiles in both the stream-wise $x$ (a) and span-wise $z$ (b) directions for $c$ and $s$ at $t=0.16$ s (after one period) are compared.
    }
    \label{fig:comparison_average_harmonic}
\end{figure}

Equation~\eqref{eq:Da_approx} employs an isotropic but harmonically averaged form of the diffusion coefficient. From Eq.~\eqref{Dh} or \eqref{Da_2}, one may alternatively approximate the diffusion coefficient using its arithmetic average as $D=\psi_1D_1+\psi_2D_2$. However, this arithmetic average completely neglects the jump in the diffusion coefficient across the interface. To evaluate the impact of this approximation, we additionally investigated how these different averaging approaches influence the results for the numerical example presented in Section~\ref{sec:moving_interface}. The analysis confirmed that the harmonic average provides higher accuracy than the arithmetic average (Fig.~\ref{fig:comparison_average_harmonic}). Therefore, although Eq.~\eqref{eq:Da_approx} does not incorporate anisotropy, it appropriately reflects the interfacial conditions in the radial direction.

\end{appendix}

\clearpage

\bibliographystyle{elsarticle-num}
\bibliography{mybibfile_marked}

\end{document}